\documentclass[twocolumn,english,aps,prx,notitlepage,superscriptaddress]{revtex4-2}

\usepackage[T1]{fontenc}
\usepackage[latin9]{inputenc}
\usepackage{graphicx}
\usepackage{amsmath}
\usepackage{float}
\usepackage{sansmath}
\usepackage{physics}
\usepackage{bm}
\usepackage{amssymb}
\usepackage[colorlinks = true,
linkcolor = blue,
urlcolor  = blue,
citecolor = blue,
anchorcolor = blue]{hyperref}
\usepackage[usenames, dvipsnames]{color}
\usepackage{changes}

\begin{document}

\sloppy
\title{Quantum Fisher Information in semiclassical magnon systems}

\author{Wolfgang Simeth}
\email{wsimeth@lanl.gov}
\affiliation{MPA-Q, Los Alamos National Laboratory, Los Alamos, New Mexico 87545, USA}

\author{Pontus Laurell}
\email{plaurell@missouri.edu}
\affiliation{Department of Physics and Astronomy, University of Missouri, Columbia, Missouri 65211, USA}
\affiliation{Materials Science and Engineering Institute, University of Missouri, Columbia, Missouri 65211, USA}

\author{Allen Scheie}
\email{scheieao@gcc.edu} 
\affiliation{MPA-Q, Los Alamos National Laboratory, Los Alamos, New Mexico 87545, USA}
\affiliation{Department of Physics, Grove City College, Grove City, PA 16127, USA}

\date{\today}

\begin{abstract}
Quantum Fisher Information (QFI) is a powerful spectroscopic tool to witness many-body quantum entanglement in solid state materials, but it is not always obvious how to relate it to other characteristics of condensed matter systems. 
In this study we elaborate on the meaning and interpretation of QFI in condensed matter by examining simple theoretical spin systems. 
We use finite sized spin systems to illustrate that QFI quantifies the momentum-dependent degree of quantum entanglement (that is the entanglement depth at a given wave vector) within a wave function. 
We subsequently use semiclassical frustrated spin models to show that, in the context of linear spin wave theory (LSWT), QFI quantifies the momentum-dependent degree of magnon squeezing in the ground state. 
In antiferromagnets with a zero-energy Goldstone mode, LSWT breaks down and the resulting semiclassical QFI (and entanglement depth) diverges at the magnetic ordering wave vector.
We also show examples of emergent quantum phases in frustrated spin systems that do not appear in classical phase diagrams. When these emergent phases are approached in parameter space, QFI calculated in the context of LSWT diverges across multiple wave vectors in momentum space. 
Taken together, QFI is not only helpful as a lower bound for entanglement depth, but may serve as a momentum-resolved probe of entanglement that offers a novel perspective on quantum critical phenomena.
\end{abstract}

\maketitle

\section{Introduction}

Quantum entanglement denotes when quantum mechanical degrees of freedom cannot be described independently. Initially introduced in the context of thought experiments ~\cite{1935_Einstein_PhysRev,1935_Schrodinger_MathProcCambPhilosSoc,bell1964einstein}, the concept of quantum entanglement has prompted longstanding philosophical controversies~\cite{1972_Freedman_PhysRevLett,2013_Yin_PhysRevLett,2004_Bell_} and is nowadays the cornerstone for applications like quantum computing and quantum sensing~\cite{Horodecki2009review, 2022_Yuan_PhysicsReports, Hung2024SensingReview}. 
Typically, entanglement is studied in the context of small ($<100$) numbers of particles. However, studying the entanglement of quantum materials across different phases and upon external perturbations may provide valuable new insights on the nature of phase transitions and quantum critical phenomena in bulk materials, including heavy fermions and high-temperature superconductors~\cite{keimerPhysicsQuantumMaterials2017}.

Recently, neutron spectroscopy has proved a powerful technique to witness quantum spin entanglement in bulk materials \cite{2025_Laurell_AdvQuantumTechnol,scheie2025tutorial}. The most robust witness so far is the \textit{Quantum Fisher Information} (QFI), which provides a lower bound for the entanglement depth \cite{Hauke2016}. It has been experimentally demonstrated in spin chains~\cite{2021_Scheie_PhysRevBa,2023_Scheie_PhysRevBa,2021_Laurell_PhysRevLett,2025_Laurell_AdvQuantumTechnol,2025_Kish_NatCommun}, in two-leg spin ladders~\cite{2024_Hong_}, in heavy-fermion materials~\cite{2025_Fang_NatCommun,2026_Mazza_}, and in triangular lattice antiferromagnets~\cite{2024_Scheie_NatPhys, scheie2026spectrum}.

Although QFI is straightforward to calculate from both experimental and theoretical spectra, it is not always obvious how to interpret the meaning of a particular entanglement depth for condensed matter. (Although condensed matter phases are often described and sometimes defined in terms of experimentally inaccessible entanglement entropies, they are rarely---if ever---characterized by their entanglement \emph{depth}.) Furthermore, there are unexplored aspects of QFI, like the wavevector dependence, that have so far not been related to entanglement depth but still carry information related to quantum coherence \cite{PhysRevResearch.6.033183} and potentially also to quantum entanglement. Still further, recent studies report that semiclassical models can have large values of QFI \cite{scheie2025tutorial}. This raises the pressing question: how to interpret QFI results to diagnose emergent quantum phases?  

In this study we elaborate on QFI in condensed matter by considering a series of simple theoretical systems. (Our aim is to use simple models to shed light on QFI rather than QFI to shed light on the models.) We begin with finite spin systems to illustrate how QFI behaves when the degree of collective quantum superposition (that is the superposition that leads to entanglement) increases, subsequently we consider the finite-size scaling properties of one-dimensional spin chains, and finally compute the QFI of semiclassical spin models using linear spin-wave theory (LSWT) to demonstrate that QFI divergences can be understood via the Bogoliubov transformation as strongly squeezed ground states. 
We demonstrate a relationship between QFI divergences and quasiparticle breakdown and illustrate that, in situations where an \textit{emergent quantum phase} appears, QFI (as evaluated through LSWT) divergences occur over multiple wave-vectors in momentum space. (In the context of this paper, ``\textit{emergent quantum phase}'' means a distinct phase in a quantum phase diagram that appears at a classical phase boundary and does not exist in the classical phase diagram.) Specifically, we investigate two- and three-dimensional examples where frustration leads to the suppression of long-range magnetic order and likely leads to quantum spin liquid or valence bond solids. In all of these cases, \emph{the semiclassical} QFI diverges on an entire line in momentum space. Momentum-dependent semiclassical QFI may be thus used as a harbinger of emergent quantum phases. 
Taken together, our study provides clear guidance on interpreting QFI in condensed matter and opens new research directions in studying the entanglement of quantum states in the thermodynamic limit. 

\section{Finite-size spin chains}
\label{Sec1}

To begin, let us briefly review the definitions of QFI. 
For a pure state $\ket{\Psi}$, the QFI density with respect to an observable $\mathcal{O}$ is given by \cite{Hauke2016}:
\begin{align}
    f_{\mathcal{Q}}:= \frac{4}{N}\cdot (\bra{\Psi}\mathcal{O}^{\dagger}\mathcal{O}\ket{\Psi}-\bra{\Psi}\mathcal{O}\ket{\Psi}^2) \, ,
    \label{eq:QFIdef}
\end{align}

where $N$ denotes the number of spins in the system. That is, $f_\mathcal{Q}$ is an operator-dependent quantity. In the context of spin entanglement, we are interested in observables $\mathcal{O}_{\bf{Q}}=S_{\alpha}(\bf{Q})=\sum_j S_j^{\alpha}\exp(\mathrm{i}\bf{Q}\cdot\bf{r}_j)$. At $T=0$ this can be written as the energy-integrated inelastic magnetic scattering $S_{\alpha\alpha}(\bm{Q},E)$, via the relation:
\begin{align}
    f_{\mathcal{Q}}[\bm{Q}] = 4 \int_{0^+}^{\infty} \mathrm{d}E  \, S_{\alpha\alpha}(\bm{Q},E)   \label{eq:Hauke}
\end{align} \cite{2023_Menon_PRB,Hauke2016} (at $T>0$ the integrand is multiplied by a hyperbolic tangent). The repeated index $\alpha$ is not summed over.
The normalized QFI (nQFI) is given by 
\begin{equation}
    {\rm nQFI}[\bm{Q}] = \frac{f_{\mathcal{Q}}[\bm{Q}]}{4S^2} \label{eq:nQFI}
\end{equation} 
and provides--for each wave vector $\bf{Q}$--a lower bound on the entanglement-depth in the spin system. If ${\rm nQFI}> m$, where $m$ is a non-negative integer, the spin system is at least $m+1$ partite entangled \cite{Hyllus_2012, Toth_2012}. As we are considering finite-sized spin system, $m$ must be a divisor of $N$ for this bound to hold. (If $m$ is not a divisor, the bound can be written  ${\rm nQFI}> sm^2+r^2,$ where $s=\lfloor N/m\rfloor$, $r=N-sm$ and $\lfloor x\rfloor$ is the floor function \cite{Hyllus_2012, Hauke2016}. We will need this version of the bound when discussing states of three spins.)

In this section, we are concerned with QFI evaluated in fully quantum-mechanical, wave function-based treatments. In later sections, we will discuss QFI as evaluated in semiclassical treatments, primarily using LSWT.
The link between entanglement and correlation functions via Eq.~\eqref{eq:nQFI} naturally also raises the question: how does the formula translate into the fully classical limit? In the limit $\hbar\rightarrow 0$, $S\rightarrow\infty$, frequency can no longer be linked to energy and the integration range in Eq.~\eqref{eq:Hauke} collapses to a single point for spin systems with bounded excitation bandwidth. Therefore, nQFI calculated from $S_{\alpha\alpha}$ vanishes. A detailed account on the classical limit is found in Ref.~\cite{2023_Scheie_PhysRevBa}.

\subsection{QFI as a measure of collective superposition}

Now let us consider some examples to build intuition about the meaning of QFI. 
Table \ref{tab:TrivialExamples} considers several simple wavefunctions and their associated nQFI from Eqs. \ref{eq:QFIdef} and \ref{eq:nQFI}. 

The most trivial example of a free spin has  entanglement depth 1 and for operator $S^z$ yields nQFI$\leq1$. 
We also consider a ferromagnet along $z$, which is not entangled in the $S_z$ basis (it is a product state). Accordingly, nQFI does not witness any entanglement and is zero. 
Meanwhile, the three-spin Greenberger-Horne-Zeilinger (GHZ) state \cite{2000_Duer_PRA} has entanglement depth 3, and nQFI with respect to $\mathcal{O}=\sum_j S_j^z$ is exactly equal to 3. We can extend this analysis to a Schr\"odinger's cat state (a superposition of two opposing states, hereafter ``cat state'') of arbitrary size, and find that for $N$ spins in a state of superposition, the appropriate operator $\mathcal{O}$ will witness the full entanglement depth ${\rm nQFI}=N$.

\begin{table}
	\centering
	\begin{ruledtabular}
		\begin{tabular}{llc}
			wavefunction   & operator & nQFI \\ \hline
			Free spin  \quad $ \frac{1}{\sqrt{a^2+b^2}} \big(a\cdot |\uparrow\rangle  +  b\cdot|\downarrow\rangle\big)$ & $\mathcal{O} = S^z$   &  $\leq1$ \\
			Uniform \quad \quad  $|\uparrow\uparrow\uparrow... \uparrow\rangle$ & $\mathcal{O} = \sum_j S^z_j$  &  0 \\
      	GHZ  \quad  \quad $\frac{1}{\sqrt{2}} \big( |\uparrow\uparrow\uparrow\rangle  -  |\downarrow \downarrow\downarrow\rangle\big)$ & $\mathcal{O} = \sum_j S^z_j$   &  3 \\
		Cat state  $\frac{1}{\sqrt{2}} \big( |\uparrow\downarrow...\rangle_N  +  |\downarrow \uparrow... \rangle_N\big)$ & $\mathcal{O} = \sum_j S^z_j e^{\mathrm{i} \pi r_j}$   &  $N$ \\
        W state $\frac{1}{\sqrt{3}} \big( |\downarrow\downarrow\uparrow\rangle + |\downarrow\uparrow\downarrow\rangle + |\uparrow\downarrow\downarrow\rangle\big)$ &    $\mathcal{O}=\sum_j \hat{n}\cdot \mathbf{S}_j$   &   $\leq \frac{7}{3}$
		\end{tabular}
	\end{ruledtabular}
	\caption{Different wave functions and their QFI properties. In the first example, $a$ and $b$ denote real or complex numbers. In the last example, $\hat{n}$ denotes a unit vector.}
	\label{tab:TrivialExamples}
\end{table}

This exercise illustrates the property of nQFI witnessing \textit{the degree of collective quantum fluctuations}. The GHZ and cat states are both quantum mechanical superpositions of two distinct spin-configuration realizations (i.e., basis states). In such cases, nQFI quantifies the number of spins that attain a different state in the two realizations, in the sense that these differences contribute to the variance of $\mathcal{O}$. 
To further illustrate this, we compute in Table~\ref{tab:CatStates} the nQFI of an eight-spin chain. We consider wavefunctions that are superposition of two possible spin-configuration realizations with an increasing number of spins that attain opposite states in these two realizations.
For this class of states, nQFI monotonically increases with the number of opposing spins in $|\psi\rangle$ that are in mutually entangled superposition. This follows from the QFI definition in Eq.~\eqref{eq:QFIdef}, where QFI is directly proportional to the ground state variance.

More generally, nQFI is not determined by the number of basis states that enter the superposition, but by how correlations in the superposition state contribute to the variance of $\mathcal{O}$. The W state $\frac{1}{\sqrt{3}} \big( |\downarrow\downarrow\uparrow\rangle + |\downarrow\uparrow\downarrow\rangle + |\uparrow\downarrow\downarrow\rangle\big)$ \cite{2000_Duer_PRA} exemplifies this. Although it is a superposition of three basis states, each term differs only locally from the others. nQFI with respect to $\mathcal{O}=\sum_j S_j^z$ is in fact zero, while the maximal value $\frac{7}{3}$ (witnessing entanglement depth $3$) can be obtained with $\mathcal{O}=\sum_j S_j^x$ \cite{Ozaydin2014}. We also note that the W state is more robust to local perturbations and measurements than the GHZ state \cite{2000_Duer_PRA}.

 \begin{table}
	\centering
	\begin{ruledtabular}
		\begin{tabular}{ll}
$ \psi =   \ket{\uparrow\uparrow\uparrow\uparrow\uparrow\uparrow\uparrow\uparrow} $  &  ${\rm nQFI}[Q = \pi] =0.0$ \\
$ \psi = \frac{1}{\sqrt{2}} \big( |\uparrow\uparrow\uparrow\uparrow\uparrow\uparrow\uparrow\uparrow\rangle  +  |\downarrow\uparrow\uparrow\uparrow\uparrow\uparrow\uparrow\uparrow\rangle\big)$  &  ${\rm nQFI}[Q = \pi] =0.125$ \\
$ \psi = \frac{1}{\sqrt{2}} \big( |\uparrow\downarrow\uparrow\uparrow\uparrow\uparrow\uparrow\uparrow\rangle  +  |\downarrow\uparrow\uparrow\uparrow\uparrow\uparrow\uparrow\uparrow\rangle\big)$  &  ${\rm nQFI}[Q = \pi] =0.5$ \\
$ \psi = \frac{1}{\sqrt{2}} \big( |\uparrow\downarrow\uparrow\uparrow\uparrow\uparrow\uparrow\uparrow\rangle  +  |\downarrow\uparrow\downarrow\uparrow\uparrow\uparrow\uparrow\uparrow\rangle\big)$  &  ${\rm nQFI}[Q = \pi] =1.125$ \\
$ \psi = \frac{1}{\sqrt{2}} \big( |\uparrow\downarrow\uparrow\downarrow\uparrow\uparrow\uparrow\uparrow\rangle  +  |\downarrow\uparrow\downarrow\uparrow\uparrow\uparrow\uparrow\uparrow\rangle\big)$  &  ${\rm nQFI}[Q = \pi] =2.0$ \\
$ \psi = \frac{1}{\sqrt{2}} \big( |\uparrow\downarrow\uparrow\downarrow\uparrow\uparrow\uparrow\uparrow\rangle  +  |\downarrow\uparrow\downarrow\uparrow\downarrow\uparrow\uparrow\uparrow\rangle\big)$  &  ${\rm nQFI}[Q = \pi] =3.125$ \\
$ \psi = \frac{1}{\sqrt{2}} \big( |\uparrow\downarrow\uparrow\downarrow\uparrow\downarrow\uparrow\uparrow\rangle  +  |\downarrow\uparrow\downarrow\uparrow\downarrow\uparrow\uparrow\uparrow\rangle\big)$  &  ${\rm nQFI}[Q = \pi] =4.5$ \\
$ \psi = \frac{1}{\sqrt{2}} \big( |\uparrow\downarrow\uparrow\downarrow\uparrow\downarrow\uparrow\uparrow\rangle  +  |\downarrow\uparrow\downarrow\uparrow\downarrow\uparrow\downarrow\uparrow\rangle\big)$  &  ${\rm nQFI}[Q = \pi] =6.125$ \\
$ \psi = \frac{1}{\sqrt{2}} \big( |\uparrow\downarrow\uparrow\downarrow\uparrow\downarrow\uparrow\downarrow\rangle  +  |\downarrow\uparrow\downarrow\uparrow\downarrow\uparrow\downarrow\uparrow\rangle\big)$  &  ${\rm nQFI}[Q = \pi] =8.0$ \\
		\end{tabular}
	\end{ruledtabular}
	\caption{N\'eel cat states growing in size along a $N=8$ spin chain with the operator $\mathcal{O}=\sum_j S^z_j e^{i Q r_j}$.}
	\label{tab:CatStates}
\end{table}

\subsection{Momentum-dependence of QFI}

By Eq. \eqref{eq:Hauke}, nQFI depends on the chosen momentum transfer $\bm{Q}$. How does one interpret this momentum dependence? This is a deep question, but here we provide three answers: 

\subsubsection{Maximal nQFI as a bound of entanglement depth}

\begin{figure}
	\centering\includegraphics[width=0.49\textwidth]{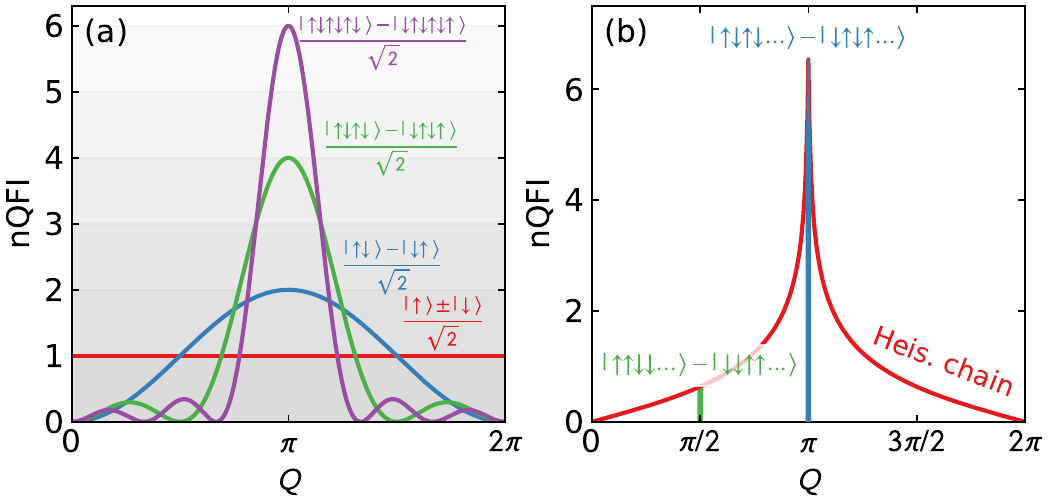}
	
	\caption{(a) Momentum-dependent QFI for spin chains of length $L=1$, 2, 4, and 6 and an antiferromagnetic cat ground state (see text). In each case the maximal ${\rm nQFI}[Q]=N$ is obtained only if $Q=\pi$. 
    (b) QFI for the 1D Heisenberg chain Bethe Ansatz solution \cite{2009_Caux_JMathPhys}. nQFI is maximal at $Q=\pi$, but nonzero over a range of $Q$ values including e.g. $Q=\pi/2$.
}
\label{fig:Figure1-chains}
\end{figure}

To give the most stringent bound on entanglement depth, one must choose the momentum transfer $\bm{Q}$ where nQFI is maximal. 
This is illustrated in Fig. \ref{fig:Figure1-chains}(a), where we compute the momentum dependent $T=0$ nQFI for small (length $N=1$, 2, 4, and 6 spin) Ising spin chains with Hamiltonian $\mathcal{H} = \sum_i (S^z_i S^z_{i+1} + \Delta S^x_i S^x_{i+1})$ where $0 < \Delta \ll 1$, and nQFI is computed along the $S^z$ basis. (The effect of the small $S^x$ exchange is to stabilize a cat state as the ground state.)  Because the ground state has all spins mutually entangled, the entanglement depth is $N$. nQFI witnesses only the full entanglement depth (${\rm nQFI}=N$), though, if we choose the wavevector ${Q}=\pi$. At several multiples of $Q=\frac{\pi}{N}$ nQFI even vanishes.
This generalizes to a system of arbitrary $N$. As $N$ grows, the QFI peak around $Q=\pi$ gets sharper so that as $N \rightarrow \infty$ the peak becomes a delta function, reflecting that, for an antiferromagnetic (AFM) $S=\frac{1}{2}$ Ising chain in the thermodynamic limit, $S(Q,\omega)\sim \delta(\omega)\cdot \delta(Q-\pi)$. With the wrong wavevector, however, one will not witness the entanglement that is there. To get the strongest bound on entanglement depth, one must choose $\bm{Q}$ to be where nQFI is maximal. (Depending on the situation, this may or may not be the most informative bound. One can also focus on wave vectors where nQFI changes the most as a control parameter is varied.)

\subsubsection{Degree of entanglement at multiple periodicities} 

Figure \ref{fig:Figure1-chains}(a) shows nonzero QFI at values other than $Q=\pi$ for one-dimensional (1D) Ising chains. However, for $N \rightarrow \infty$ the QFI at values other than $Q=\pi$ is suppressed (see above). 

A generic quantum ground state, however, is not a cat state and can involve superposition of more than two spin configurations. This is illustrated in Fig. \ref{fig:Figure1-chains}(b) which plots the $T=0$ nQFI calculated from the algebraic Bethe Ansatz solution of the 1D Heisenberg chain \cite{2009_Caux_JMathPhys}. 
The largest nQFI is found at $Q=\pi$, in accord with a quantum ground state dominated by N\'eel correlations (and for a Hamiltonian for which staggered magnetization is a relevant operator in the renormalization group sense). However, momenta away from $Q=\pi$ also have nonzero nQFI---which indicates entanglement of spin configurations in the ground state, where collective superpositions involve periodicities beyond the $\pi$-modulation of a cat state. 
Unlike the Bethe ansatz solution, cat states in the 1D Heisenberg chain decohere upon infinitesimal perturbation \cite{Chen_2021_QCM}. (Tnhis raises the question if wave functions for which nQFI is accumulated at a single wave vector can form stable entangled ground states in the thermodynamic limit---the answer may be ``no''.)
Thus, QFI is more than just a bound on entanglement depth: its $Q$-dependence reveals the degree of collective quantum superposition at various wavevectors. As we will argue below, it may further be used to distinguish entangled antiferromagnets from emergent quantum phases, where frustration results in the breakdown of long-range magnetic order~\cite{zhou2026quantum}. 

\subsubsection{Quantum coherence length}

Ref.~\cite{PhysRevResearch.6.033183} calculated the ``quantum coherence length'', originally introduced in Ref.~\cite{2016_Malpetti_PhysRevLett}, from the spatial decay of the QFI matrix (QFIM). Equivalently, this is represented by the sharpness of the features in ${\rm nQFI}[\bm{Q}]$: a sharp peak indicates long-ranged entanglement; a broad peak indicates short-ranged entanglement (collective superposition only with neighboring spins).
We say no more about it here, but this is a third and important interpretation of of the $\bm{Q}$-dependence of QFI: the spatial extent of quantum correlations.

It is, however, instructive to contrast the quantum coherence length with the more conventional total correlation length. The total correlation length can be calculated for spin components either \cite{forster1975hydrodynamic} from the disconnected correlation function
\begin{equation}
    C_{ij,\mathrm{d}}^{\alpha\beta} = \langle S_i^\alpha S_j^\beta\rangle,
\end{equation}
or from the connected correlation function
\begin{equation}
    C_{ij,\mathrm{c}}^{\alpha\beta} = \langle S_i^\alpha S_j^\beta\rangle - \langle S_i^\alpha \rangle \langle S_j^\beta \rangle.
\end{equation}
Distinguishing between the two becomes important in symmetry-broken states (and most systems considered in this work have spontaneously broken symmetries). The disconnected piece gives rise to Bragg peaks and its decay is used to define the standard correlation length in neutron scattering, which is helpful for characterizing magnetically ordered phases \cite{boothroyd2020principles}. The connected correlation function describes fluctuations \cite{forster1975hydrodynamic}. The total correlation length calculated from $C_{ij,\mathrm{c}}^{\alpha\beta}$ and the quantum coherence length coincide (up to constant prefactors) at $T=0$ but generically disagree at $T>0$.

\subsection{Scaling properties of QFI}

The QFI scaling with the system size $N$ is highly dependent on the ground state (e.g. whether it's a cat state, a GHZ state, etc.). The maximal rate at which it can grow in equilibrium is known as the Heisenberg limit, where $f_{\bf{Q}}\sim N$, and is realized by cat states. (This has been elaborated by quantum metrology~\cite{2025_Laurell_AdvQuantumTechnol,1969_Helstrom_JStatPhys,2011_Holevo_,Toth_2014}.) While this growth rate exists in theory, in general it can be limited by perturbations, including by noise or couplings to thermal baths \cite{Ozaydin2014, Toth_2014, Chen_2021_QCM, 2023_Baykusheva_PhysRevLett}. Here, we focus on the zero-temperature scaling of closed systems, with the understanding that this is a theoretical upper bound. In this context, Ref.~\cite{Hauke2016} showed that the transverse field Ising model displays ${\rm QFI} \sim N^{3/4}$.

We compute the scaling of the finite-sized 1D Heisenberg model using Lanczos exact diagonalization \cite{1950_Lanczos} and density matrix renormalization group (DMRG) 
calculations for system sizes up to $N=1024$. Results, as obtained for periodic and open boundary conditions are discussed in Appendix~\ref{Appendix4}. 
In accord with the numerical results from Hallberg et al.~\cite{1995_Hallberg_PhysRevB}, we find
\begin{align}
    \mathrm{nQFI}(Q=\pi)\sim \left[ \log\left(\frac{cN}{2}\right)  \right]^{3/2}  \, ,  \label{eq:HAFM:scalingform}
\end{align}
for $c=\text{const}$. 
This scaling, which is weaker than an entanglement depth ``volume-law'', is reminiscent of the known log-scaling of the entanglement entropy in critical 1D systems~\cite{2010_Cardy_JStatMech,Vidal_2003,Vitagliano_2010} (though this terminology is only precisely defined for entanglement entropy and not QFI \cite{RevModPhys.82.277}). The found log-form does not follow from the scaling analysis in Ref.~\cite{Hauke2016}, which predicts power-law scaling at many highly entangled quantum critical points. This apparent deviation likely arises because correlation functions of the quantum critical 1D Heisenberg model are known to have significant logarithmic corrections, which were not included in the scaling analysis of Ref.~\cite{Hauke2016}.

The QFI scaling with system size is not an easily-accessible experimental quantity, but from a theoretical perspective a positive scaling with $N$ signals systems that can, in principle, reach large entanglement depths. The fact that the thermodynamically stable 1D Heisenberg chain scales logarithmically shows that it indeed follows a sub-volume law QFI scaling as opposed to a cat state, which displays volume law QFI scaling but which is unstable in the thermodynamic limit. Studying this scaling may therefore be crucial for identifying quantum states that feature high entanglement depth, while being thermodynamically stable, being robust against decoherence, and surviving interaction with their environments.

\section{QFI in semiclassical models}
\label{Sec2}

One might expect that a semiclassical model of non-interacting magnons (i.e. linear spin wave theory) would produce an unentangled ground state---but this is not generically the case. Indeed, linear spin wave theory is expressed in terms of quadratic bosonic Hamiltonians, and can thus be viewed as a quantum description of coherent quasiparticles. Although these arise as corrections above classical product ground states, the magnon vacuum is not necessarily trivial \cite{2019_Kamra_PhysRevB,2022_Wuhrer_PhysRevB,2025_Rozsa_PRB}. In particular, gapless antiferromagnets are in fact highly entangled states of matter which can have diverging semiclassical QFI (see Fig.~\ref{fig:Figure1b}). Here we demonstrate this with simple non-interacting magnon models. We generically expect these divergences to be tempered in a fully quantum-mechanical treatment. Nevertheless, they appear to herald nearby quantum phases.

\begin{figure*}
    \includegraphics[]{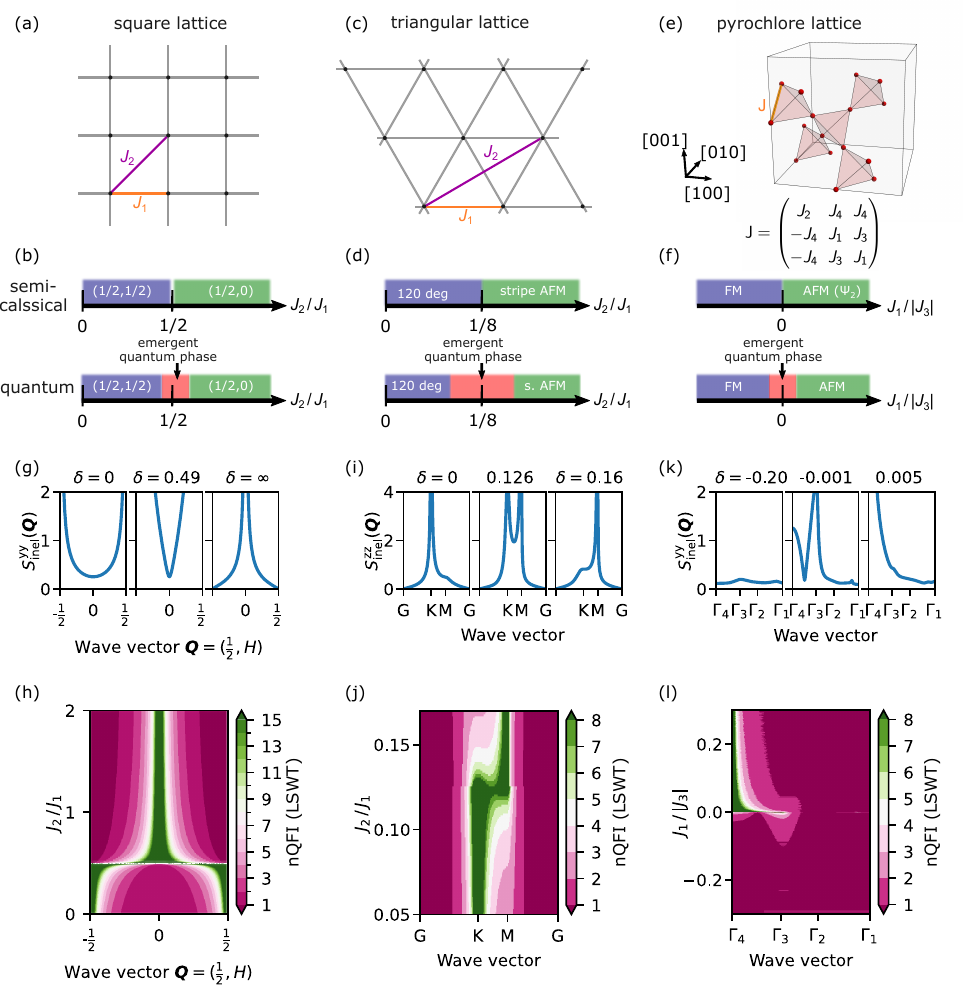} 
    \caption{Semiclassical QFI in prototypical frustrated spin systems with quantum phase transitions. (a) Square lattice antiferromagnet with nearest and next-nearest neighbor Heisenberg interactions ($J_1>0$ and $J_2>0$) along the square-lattice edge (orange) and square lattice diagonal (purple), respectively. The nodes (black circles) are decorated with spins ($S=\frac{1}{2}$). (c) Triangular lattice antiferromagnet of $\frac{1}{2}$-spins (black circles) with Heisenberg interactions along nearest neighbor ($J_1>0$, orange) and along next-nearest neighbor ($J_2>0$, purple) bonds. (e) Pyrochlore lattice. The red points denote $S=\frac{1}{2}$ spins decorating a pyrochlore lattice characterized by point sharing tetrahedra and are generated by the Wyckoff position $b$ in space group Fd$\bar{3}m$ (number 227). The exchange matrix $\mathrm{J}$ is given along nearest-neighbor bonds and described in the main text.
    (b,d,f) Semiclassical and quantum phase diagrams for the three models shown in (a,c,e), respectively. 
    In the square lattice, (b), antiferromagnetic order with wave-vector $\bm{k}=(\frac{1}{2},\frac{1}{2})$ is stabilized for dominant next-nearest neighbor exchange ($J_2/J_1\ll\frac{1}{2}$) and with wave-vector $\bm{k}=(\frac{1}{2},0)$ for dominant next-nearest neighbor exchange ($J_2/J_1\ll\frac{1}{2}$), respectively. In (d) antiferromagnetic $120^{\circ}$ order is seen for small values of $J_2/J_1$ and stripe AFM order for large values of $J_2/J_1$. In the pyrochlore lattice [panel (f)] ferromagnetic and antiferromagnetic order are stabilized for small and large values of $J_1/|J_3|$, respectively.
    In the semiclassical cases, these sets of two ordered phases are separated by the points $\frac{1}{2}$, $\frac{1}{8}$, and 0, respectively, whereas in the quantum phase diagram they are separated by a quantum phase, where static magnetic order breaks down.  
    (g), (i), and (k) Dynamical spin structure factor for different values of $\delta$ calculated using LSWT for the three models shown in (a), (c), and (e), respectively and the semiclassical phases shown in (b), (d), and (f), respectively. Points K and M in the reciprocal space of the triangular lattice denote $(\frac{1}{3},\frac{1}{3})$ and $(\frac{1}{2},0)$, respectively. The points $\Gamma_4$, $\Gamma_3$, $\Gamma_2$, and $\Gamma_1$ denote the points $(202)$, $(200)$, $(002)$, and $(000)$, respectively. (h), (j), and (l) semiclassical nQFI as a function of $\delta$ and momentum transfer for the three models that are shown in (a),(c), and (e), respectively, calculated from the LSWT dynamical spin structure factors.
    }    
    \label{fig:Figure2}
\end{figure*}

\subsection{Square lattice }

Let us consider the square lattice antiferromagnet with first and second neighbor exchanges $J_1$, $J_2$ and single-ion anisotropy $\kappa$
\begin{equation}
H =  J_1 \sum_{\left\langle i,j\right\rangle}  \bm{S}_{i}\cdot\bm{S}_{j} + J_2 \sum_{\left\langle\langle i,j\right\rangle\rangle}  \bm{S}_{i}\cdot\bm{S}_{j} + \kappa \sum_i (S_i^z)^2  \, ,
\label{Eq:SquareHamiltonian}
\end{equation}
where $\langle i,j\rangle$ and $\langle \langle i,j\rangle \rangle$ denote sums over the first and second neighbor bonds, respectively. 
The classical phase diagram with $J_1>0$ and $J_2>0$ and $\kappa=0$ has a phase boundary at $\delta:=J_2\,/\,J_1=\frac{1}{2}$ [Fig. \ref{fig:Figure2}(b)] between $(\frac{1}{2},\frac{1}{2})$ order for $\delta\,\,<\frac{1}{2}$ and $(\frac{1}{2},0)$ order for $\delta\,\,>\frac{1}{2}$ (shown in Fig. \ref{fig:FigureGroundStates}) \cite{2009_Murg_PhysRevB}. 
In the quantum limit, this phase boundary becomes a quantum phase with no static magnetic order extending asymmetrically around the classical critical point~\cite{1988_Chandra_PhysRevB,1988_Ioffe_IntJModPhysB,Dagotto_1989,1993_Ferrer_PhysRevB} (though it is debated whether this phase is a spin-liquid or valence bond solid \cite{4yrt-nsth,Haghshenas_2018,2024_Qian_PhysRevB,2018_Wang_PhysRevLett}).

We use linear spin-wave theory, described analytically in Appendix \ref{Appendix1}, to calculate the semiclassical dynamical structure factor $S^{yy}(\bm{Q},\omega)$. Figure \ref{fig:Figure2}(g) shows the inelastic part, $S^{yy}_{\mathrm{inel}}(\bm{Q}):=\int_{0+}^{\infty}\mathrm{d}\omega\,S^{yy}(\bm{Q},\omega)$, for different values of $\delta$ on the momentum space line $\bm{Q}=(\frac{1}{2},H)$. For dominant nearest-neighbor exchange ($\delta=0$), spectral weight diverges around $\bm{Q}=(\frac{1}{2},\frac{1}{2})$. In the opposite limit of dominant next-nearest neighbor interaction, intensity is likewise strongest at $\bm{Q}=(0,\frac{1}{2})$. Closer to the critical point $\delta=1/2$, spectral weight is enhanced on the entire path shown, and diverges to infinity when $J_1 = 2 \> J_2$. 

This $S({\bf Q},\omega) \sim 1/\omega$ intensity divergence is a well-known feature of gapless antiferromagnets, and corresponds to an instability where the Bogoliubov transform also diverges. By Eq.~\eqref{eq:Hauke} this means the QFI and therefore entanglement depth diverges to infinity as well. This is plotted in Fig.~\ref{fig:Figure2}. 
In the two antiferromagnetic phases of the square lattice AFM, LSWT yields $S(\boldsymbol{Q},\omega)\sim S$ for a given wave vector $\boldsymbol{Q}$.
In the context of LSWT, the classical limit, $S\rightarrow\infty$, therefore produces vanishing entanglement depth for a given momentum transfer away from the magnetic ordering wave vector $\bm{k}$, i.e., $\lim_{S\rightarrow\infty, \bm{Q}\neq k} \mathrm{nQFI}[\bm{Q}]\sim 1/S$.

This QFI divergence means that the square lattice antiferromagnet---a simple physical system---is actually highly entangled already in its magnetically ordered regime (at zero temperature) that is well described in the semiclassical approximation. 
Why and how this occurs is explained by the Bogoliubov transform--a hyperbolic rotation used to diagonalize the Hamiltonian and necessary when the system is not diagonal in the N\'{e}el basis. For a gapless Heisenberg antiferromagnet the Bogoliubov transform into the diagonal basis produces a squeezed magnon vacuum with diverging (semiclassical) entanglement depth. And the larger the matrix rotation angle, the bigger the off-diagonal elements, and the larger the entanglement.

Now, this is a true divergence only with a perfectly gapless Goldstone mode. A small gap opening will generically cause the divergence to become finite. The zone-center nQFI in the $(\pi\pi)$ phase for $J_2=0$ in the presence of an easy-axis anisotropy $\kappa<0$ is  
\begin{align}
    {\mathrm{nQFI}}=\frac{\sqrt{|\kappa|(|\kappa-4J_1|)}}{|\kappa|}
\end{align}
At $\kappa =0$ the nQFI is infinite, but even with a small anisotropy the computed entanglement depth can easily be in the thousands. (For easy-plane anisotropy $\kappa>0$, nQFI is still infinite).

\begin{figure}
	\centering\includegraphics[]{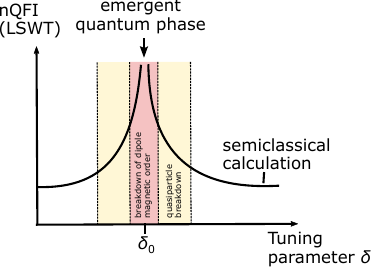}
	
	\caption{Proximity to emergent quantum phase signaled by semiclassical calculations. In the present study, we consider spin systems that are driven through an emergent quantum phase (red shading) via tuning of parameter $\delta$, where static magnetic order breaks down. The quantum Fisher information (black line), calculated with linear spin wave theory (LSWT) for a wave-vector $\bm{Q}$, displays a divergence at a critical point $\delta_0$ that lies in the region of the respective quantum phase. This is surprising, as LSWT is unable to model physics within this emergent phase. Far away from the critical point, nQFI is relatively small, consistent with trivial entanglement depth. In the intermediate region in that the quasiparticle picture gradually breaks down (yellow shading), nQFI is enhanced. 
    This characteristic $\delta$ dependence of the semiclassical nQFI is observed across multiple wave vectors for several different example models that we considered and heralds the nearby emergence of an emergent quantum phase. 
}
\label{fig:Figure1b}
\end{figure}

Of course, a diverging inelastic intensity in LSWT generically signals a breakdown of the non-interacting magnon approximation. LSWT is further based on perturbations from a classically ordered ground state---and as such it cannot make predictions about phases where classical magnetic order does not occur. From that perspective, it is intriguing that the entanglement depth (that is nQFI) diverges on a line of reciprocal space when the system approaches a quantum critical phase transition $J_2 = \frac{1}{2}J_1$ that is surrounded by an emergent phase. As a real system approaches this point, the magnons become strongly interacting and eventually break down within the quantum phase without static magnetic order (which is not captured by LSWT as it neglects magnon interactions). However, within the ``worldview'' of LSWT, a sign of a nearby emergent quantum critical phase is diverging intensity across a continuous range of wavevectors (illustrated in Fig. \ref{fig:Figure1b}). 

\subsection{Frustrated lattices}

This behavior of diverging entanglement depth near quantum critical points is not unique to the square lattice, but occurs also in several frustrated LSWT models. 
In Figure~\ref{fig:Figure2} we consider an isotropic triangular lattice antiferromagnet, and an anisotropic pyrochlore magnet. Both can be tuned through an instability where static magnetic order breaks down, which appears as a semiclassical QFI divergence along a continuous range of wavevectors.  

The $J_2/J_1$ triangular lattice model [Fig.~\ref{fig:Figure2}(d)] has a classical phase boundary at $J_2/J_1 = 1/8$ separating $120^{\circ}$ antiferromagnetic order with wave-vector $\bm{k}=(\frac{1}{3},\frac{1}{3})$ (K-point) and stripe magnetic order with wave-vector $\bm{k}=(0,\frac{1}{2})$ (M-point)~\cite{2015_Hu_PhysRevB}. In the quantum limit this point becomes an extended quantum spin liquid phase~\cite{1990_Jolicoeur_PhysRevB,1988_Huse_PhysRevLett,1973_Anderson_MaterialsResearchBulletin}. 

We also consider the anisotropic pyrochlore model with nearest neighbor exchange matrix
{\sansmath 
\begin{align}
  \mathsf{J}= \begin{pmatrix}
J_2 & J_4 & J_4\\
-J_4 & J_1 & J_3\\
-J_4 & J_3 &  J_1
\end{pmatrix}   \, ,
\end{align}}
following the conventions in Refs.~\cite{2011_Ross_PhysRevX,2017_Yan_PhysRevB}.
We consider the case where $J_4=0$, $J_2=0$, and $J_3=-1$. Tuning $\delta := J_1/|J_3|$ transforms the canted ferromagnetic (FM) order ($J_1/|J_3|<0$) to a noncoplanar antiferromagnetic order typically referred to as $\Gamma_5$ order ($J_1/|J_3|>0$) ~\cite{2017_Yan_PhysRevB}. 
Classically, there is a phase boundary at $J_1=0$ \cite{2016_Benton_NatCommun,2017_Yan_PhysRevB} which in the quantum limit becomes an extended phase with suppression of static magnetic order \cite{Gresista_2025,Zhang_YTO_2025}. 

We compute the LSWT spectra for both the triangular and pyrochlore models numerically with the \texttt{Sunny} software package~\cite{2025_Dahlbom_}, see Appendix~\ref{Appendix2}. Similar to the 2D square lattice, the semiclassical QFI diverges at the momentum of a gapless Goldstone mode within the ordered, antiferromagnetic phase. Around the quantum phase transition, this becomes a divergence along a line in reciprocal space. 
The flat zero-energy magnon may signal that the quantum phase boundary has an extensive degeneracy. As a matter of fact, we find this effect in every example that we considered where a putative emergent quantum phase appears: the 2D square lattice, the 2D triangular lattice, the 3D pyrochlore lattice, and the 2D honeycomb Kitaev model (Appendix \ref{app:Heis-Kitaev}).

\subsection{Quantum phase transitions without diverging QFI}

However, not every zero-temperature phase transition has a line-diverging nQFI. 
Two counterexamples are (i) a field-driven quantum dimer ladder, and (ii) a cubic pyrochlore ferromagnet in a magnetic field. 

\subsubsection{Field-driven quantum dimer ladder}

\begin{figure*}
    \includegraphics[]{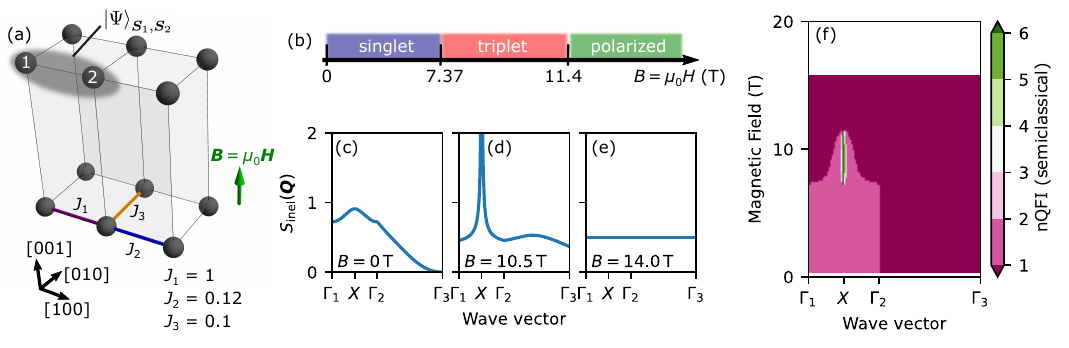} 
    \caption{Semiclassical simulations for the Heisenberg spin ladder as a function of external field. (a) Tetragonal unit cell with two atoms that carry spin-$\frac{1}{2}$. We considered the intra-chain couplings $J_1$ (purple bond) and $J_2$ (blue bond) as well as relatively weak inter-chain coupling $J_3$ (orange bond). External field was applied along the tetragonal $[001]$ axis. For semiclassical simulations accounting for singlet and triplet states, we described the pair of spins, $\bm{S}_1$ and $\bm{S}_2$, in a wave-function $\ket{\Psi}_{\bm{S}_1,\bm{S}_2}$ ($\mathrm{SU}(4)\rightarrow\mathbb{C}^2$). (b) Semiclassical phase diagram. At magnetic fields below $7.37\,$T, a singlet state is favored. In the intermediate region, a triplet state is stabilized. Above $11.5\,$T magnetic field, the state corresponds to a uniform state that is polarized along the field direction.
    (c)-(e) Dynamical spin structure factor integrated over finite energy transfers $]0,\infty[$ at external fields (c) $0\,$T, (d) $10.5\,$T, and (e) $14$\,T. The wave-vector path connects the points $\Gamma_1$, $X$, $\Gamma_2$, and $\Gamma_3$, which denote $(0,1,0)$, $(\frac{1}{2},1,0)$, $(1,1,0)$, and $(0,0,0)$, respectively.  (f) Semiclassical nQFI, along the path $\Gamma_1X\Gamma_2\Gamma_3$ as a function of magnetic field. 
    }
    \label{fig:FigureS1}
\end{figure*}

Quantum dimers in a magnetic field ($B$) are prototypical examples of quantum phase transitions~\cite{sachdev1999quantum,2014_Zapf_RevModPhysa}, where a quantum singlet at $B=0$ is separated from a long-ranged ordered antiferromagnet at finite field~\cite{2003_Ruegg_Naturea,2014_Zapf_RevModPhysa}. 
Here we consider a simple model of a spin-$\frac{1}{2}$ Heisenberg ladder in tetragonal symmetry under external magnetic field along $[001]$ [see Fig.~\ref{fig:FigureS1}(a)] with alternating intra-chain coupling, $J_1=1$ and $J_2=0.12$, as well as uniform inter-chain coupling given by $J_3=0.1$. 
The phase diagram in Fig. \ref{fig:FigureS1}(b) has field-driven transitions between singlet, triplet, and polarized phases. Unlike the previous spin models, there is no extended region in the quantum phase diagram where static magnetic order breaks down and an emergent quantum phase appears.

We calculated the LSWT spectrum using the entangled units formalism in \texttt{Sunny}, where the inter-dimer physics is treated exactly, and the intra-dimer physics is treated semiclassically \cite{Dahlbom_2024}. 
The energy-integrated spin structure factor, $S_{\mathrm{inel}}(\bm{Q})=S^{xx}_{\mathrm{inel}}(\bm{Q})+S^{yy}_{\mathrm{inel}}(\bm{Q})+S^{zz}_{\mathrm{inel}}(\bm{Q})$ is presented for several different fields in Fig. \ref{fig:FigureS1}(c), (d), and (e). 
In the singlet phase [panel (c)], spectral weight is nonuniform but finite. 
In the triplet phase [panel (d)], the intensity diverges at the X point, which corresponds to the magnetic ordering vector, but remains at relatively low values everywhere else. In the field polarized state, intensity is uniform in momentum space, as typical of a ferromagnetic state~\cite{scheie2025tutorial}. 
Normalized Quantum Fisher Information, obtained by normalizing $S_{\mathrm{inel}}(\bm{Q})$ by $4/12S^2$, is presented in Fig.~\ref{fig:FigureS1}(f). At low fields in the singlet phase, nQFI exceeds one, therefore witnessing bipartite entanglement in accord with dimer singlets \cite{scheie2025tutorial}. In the triplet phase, nQFI diverges at the Y point, similar to divergences at antiferromagnetic ordering vectors. In the field polarized state, nQFI is smaller than one, consistent with a trivial entanglement depth. 

Importantly, nowhere does nQFI diverge along a line of reciprocal space. And importantly, there is no extended region of an emergent quantum phase, where static magnetic order breaks down, at this phase boundary~\cite{2014_Zapf_RevModPhysa}. 

\subsubsection{Pyrochlore Lattice under Field}

\begin{figure*}
    \includegraphics[]{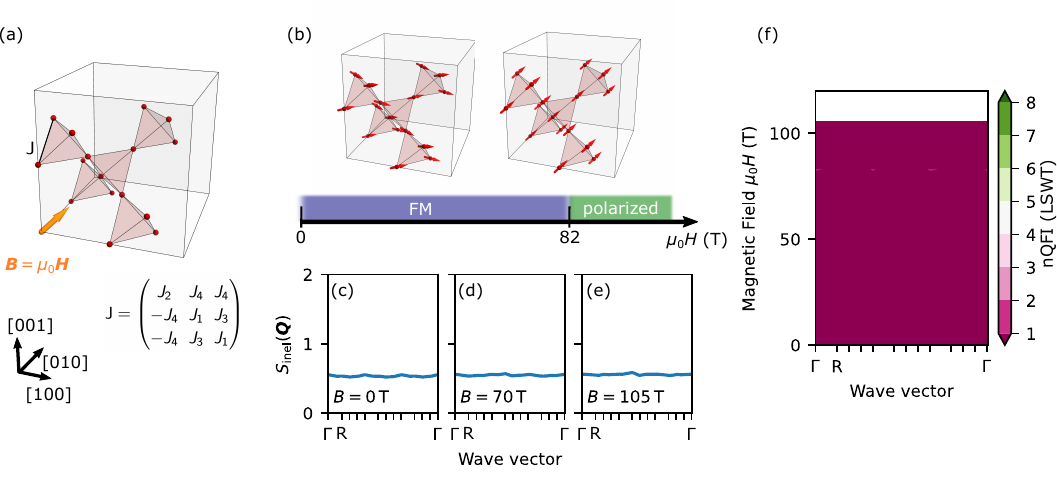} 
    \caption{Semiclassical simulations for a pyrochlore lattice as a function of external magnetic field. (a) Unit cell of the pyrochlore lattice. Along the nearest neighbor bond connecting $(\frac{1}{2},0,0)$ with $(\frac{3}{4},\frac{1}{4},0)$ we considered the exchange matrix as indicated and where $J_1=-1$, $J_3=-1$, $J_2=0$, and $J_4=0$. Magnetic field was applied along the cubic $[1,1,1]$ axis. (b) Phase diagram as a function of magnetic field. At low fields, the ground state corresponds to a canted ferromagnetic texture. At high field, a polarized state is accessed, where all spins are parallel. The two situations are illustrated schematically. (c)-(d) Dynamical spin structure factor, integrated over the dynamical range for the external field values (a) $0$, (b) $70\,$T, and (c) $105\,$T (e). Semiclassical phase diagram as a function of external field. At zero magnetic field, the system displays a canted ferromagnetic ground state. At around 4.7\,T spins become polarized uniformly along external field axis. (f) Semiclassical nQFI as a function of external field. The path on the horizontal axis connects the points (indicated by vertical ticks) $\Gamma=(0,0,0)$, $\mathrm{R}=(\frac{1}{2},\frac{1}{2},\frac{1}{2})$, $(\frac{1}{2},\frac{1}{2},0)$, $(\frac{1}{2},0,0)$, $(0,0,0)$, $(\frac{1}{2},-\frac{1}{2},\frac{1}{2})$, $(0,-\frac{1}{2},\frac{1}{2})$, $(0,-\frac{1}{2},0)$, $(0,0,0)$, $(-\frac{1}{2},-\frac{1}{2},-\frac{1}{2})$, $(-\frac{1}{2},0,-\frac{1}{2})$, $(0,0,-\frac{1}{2})$, and $\Gamma=(0,0,0)$, respectively.}
    \label{fig:FigureS3}
\end{figure*}

A cubic pyrochlore ferromagnet in a magnetic field is another system with a zero-temperature phase boundary without an emergent quantum phase where static order breaks down in an extended parameter space. We consider a pyrochlore lattice ferromagnet, which has a well-studied experimental analogue Yb$_2$Ti$_2$O$_7$ \cite{Scheie_2017_Reentrant,Saubert_2020_Orientation}. Unit cell and notations are the same as above, and magnetic field is applied along the cubic $[111]$ axis, as shown in Fig.~\ref{fig:FigureS3}(a). The semiclassical phase diagram shown in Fig.~\ref{fig:FigureS3}(b) is characterized by ferromagnetic order for fields below $H_c =82\,$T, and a polarized state above $H_c$. Energy integrated inelastic $S_{\mathrm{inel}}(\bm{Q})=S^{xx}_{\mathrm{inel}}(\bm{Q})+S^{yy}_{\mathrm{inel}}(\bm{Q})+S^{zz}_{\mathrm{inel}}(\bm{Q})$ calculated with \texttt{Sunny} are shown in Fig.~\ref{fig:FigureS3}(c), (d), and (e), respectively, and display nearly flat momentum-space dependence. 

The semiclassical nQFI  is presented in (f). In the entire field range, nQFI is below 1 and witnesses (at least) trivial entanglement depth. The lack of divergence is because the field-polarized phase boundary corresponds to a quadratic magnon band touching zero energy \cite{Scheie_2020_YTO} rather than a a linear Goldstone mode---which consequently produces no $1/\omega$ divergence in the inelastic structure factor \cite{boothroyd2020principles}. 
From another viewpoint, this phase boundary appears to be first-order \cite{Scheie_2017_Reentrant} and therefore does not display positive scaling exponent of QFI (cf. Ref.~\cite{Hauke2016}). Consequently, QFI does not increase with the system size and stays at finite values.

\section{Discussion}

The model calculations in this paper begin to give a clearer picture of the meaning of QFI in condensed matter systems. 

The finite-sized chains show that, beyond merely being a bound on entanglement depth, finite QFI across multiple wave-vectors indicates entangled, collective superposition at these wave-vectors. Specifically, higher QFI implies stronger variance of the probed observable, which can be limited by perturbations. 
The momentum-dependence in turn also carries key information: to bound entanglement depth one must choose the appropriate wavevector, but it also indicates whether the wave function involves a distribution of different basis states encoding different periodicities, and thus an extended quantum coherence length. 

The semiclassical simulations show that diverging (semiclassical) nQFI is not merely a feature of ``exotic'' phases where magnons break down, but also of conventional, semiclassical antiferromagnets. The reason is because antiferromagnetic ground states are actually highly-entangled ``squeezed states.'' 
Spin-squeezing is a multi-partite entanglement property of quantum states that is desired both for detection purposes of entanglement and in the context of quantum metrology~\cite{Toth_2014, 2022_Comparin_PhysRevLett,2025_Caleca_PhysRevLett}.
Here it is witnessed directly by QFI, which diverges to infinity as antiferromagnetic gaps go to zero (in theory at least---real experimental materials inevitably have nonzero anisotropy leading to a nonzero gap), thereby reflecting breakdown of magnon quasiparticles. The hyperbolic rotation angle of the Bogoliubov transform, which encodes the matrix rotation from the N\'{e}el ordered basis, scales with the magnitude of nQFI at the same wave-vector. Therefore, when nQFI diverges also the Bogoliubov rotation becomes degenerately strong. More broadly, this implies any magnetic system where a Bogoliubov transform is required is also entangled. 

We tuned several models through quantum critical phase boundaries that are surrounded by extended, emergent phases in the quantum phase diagram, where magnetic order breaks down.  
In all these examples, we found that the semiclassical QFI diverges to infinity along a continuous line of wavevectors (Fig. \ref{fig:Figure2}). However, not all zero-temperature phase transitions have this multiple-wavevector divergence: we observed it only at extended regions, where magnetic order breaks down and emergent quantum phases are expected.  In this way, semiclassical QFI is a harbinger of exotic (non-magnon) physics.

However, a diverging QFI in the semiclassical models does not necessarily imply diverging entanglement depth in the quantum limit since the exotic quantum phases may have finite QFI (c.f. quantum spin ice \cite{zhou2026quantum} or the 2D triangular QSL \cite{scheie2026spectrum}) with respect to the same observable. The exotic phases may still be highly entangled in a different basis or measure \cite{2026_Lyu_NatCommun}.

In the past, theoretical studies have already shown that QFI can diverge at quantum critical phase transitions (e.g., the transverse field Ising model \cite{Hauke2016}, the Kitaev chain of spinless fermions in a lattice with variable-range pairing \cite{Pezze_2017}, and the Kondo lattice \cite{2025_Fang_NatCommun}). Here we presented several examples of this behavior in insulating spin systems. 
Importantly, we showed further that also the wavevector dependence is important: a single wavevector diverging does not indicate exotic physics, but a continuum of wavevectors does. In this way, QFI is able to distinguish quantum phase transitions with emergent quantum phases from quantum phase transitions without emergent quantum phases. 

Our study does not provide a rigorous proof that QFI divergences across multiple wavevectors always accompany exotic phenomena. However, they suggest a future research direction that merits further theoretical and experimental studies and which may guide us to a novel perspective on quantum critical phenomena in the context of quantum entanglement.

\section{Conclusion}
\label{Sec5}

In summary, we studied the interpretation of quantum Fisher information in relatively simple spin systems. 
In finite-size spin systems we discussed the wave-vector dependence of nQFI, the meaning of the amplitude of nQFI, as well as the scaling of nQFI as a function of system size. 
Studying semiclassical spin models, we found that enhanced nQFI at a wave-vector translates directly to stronger quasiparticle squeezing and therefore enhanced entanglement for quasiparticles of the same wave-vector. 
Considering several magnetically ordered, frustrated systems where a quantum phase transition emerges, we found that in regions of the respective quantum phase diagrams, where static magnetic order breaks down, semiclassical simulations display strongly enhanced QFI across multiple wave-vectors potentially signaling the emergence of emergent quantum phases. 
We found this in the square-lattice antiferromagnet, the triangular lattice antiferromagnet, the pyrochlore lattice, and the honeycomb Kitaev model. However, the dimer ladder and cubic ferromagnet confirmed that not all quantum phase boundaries have diverging QFI: those without divergences appear to also lack emergent quantum phases. 
Taken together, QFI may serve as a witness of proximity to emergent quantum phases and breakdown of static magnetic order in materials such as strange-metals~\cite{2020_Cha_ProcNatlAcadSci,2019_Husain_PhysRevX,2003_Custers_Nature}, cuprates~\cite{2019_Proust_AnnuRevCondensMatterPhys,2015_Keimer_Nature,2010_Armitage_RevModPhys}, and heavy-fermions~\cite{2016_Wirth_NatRevMater,2002_Sidorov_PhysRevLetta,2007_Tanatar_Science}.

This study highlights QFI as a quantity that is much more than just a witness of multipartite entanglement. It probes entangled quantum superposition at different wavevectors. While nontrivial values of QFI alone do not necessarily indicate exotic physics, its enhancement over many momenta indicates--in all examples that we discuss-proximity to emergent quantum phases of matter where ordinary magnon physics breaks down. 
Our study therefore suggests momentum-resolved QFI accessible with neutron spectroscopy as a unique probe for exotic phenomena and highlights that quantum criticality---a long-standing research area in phyics---may gain important new insight from experimental and theoretical studies of quantum Fisher information in condensed matter.

\subsection*{Note added}

While this manuscript was in its final stages of preparation, we became aware that Fang \textit{et al.} also studied QFI of the square-lattice antiferromagnet \cite{2026_Fang_}. The two studies are complementary and in regions of overlap consistent with each other.

\acknowledgments

The work by WS and AS was supported by the Laboratory Directed Research and Development program of Los Alamos National Laboratory under project number 20250836ECR. 
We acknowledge helpful discussions with Alan Tennant, Kipton Barros, David Dahlbom, Tommaso Roscilde, and Esteban Ghioldi.
Exact diagonalization and DMRG computations for this work were performed on the high performance computing infrastructure operated by Research Support Solutions in the Division of IT at the University of Missouri, Columbia MO \footnote{\url{https://doi.org/10.32469/10355/97710}}.

%


\appendix

\section{Bogoliubov transformation and Quantum Fisher Information}
\label{Appendix1}

In the following, we demonstrate that QFI divergences in antiferromagnets can be understood semiclassically in the context of Bogoliubov transformations. For this, we consider the square lattice spin-$\frac{1}{2}$ antiferromagnet with nearest and next-nearest neighbor interaction ($J_1$ and $J_2$) as well as anisotropy given by Eq. \eqref{Eq:SquareHamiltonian} in the main text. In the following, we consider the model in the absence of anisotropy ($\kappa=0$), unless stated otherwise. 

\subsection{General procedure}
To solve the LSWT by the common procedure, as accounted for in Refs.~\cite{1961_Kaplan_PhysRev,2015_Toth_JPhysCondensMatter,2009_Chernyshev_PhysRevB}, we start with bosonic operators $\hat{a}_i$, $\hat{a}^{\dagger}_i$ associated with creation and annihilation at site $\bm{r}_i$. The operators satisfy $[\hat{a}_i,\hat{a}^{\dagger}_j]=\delta_{ij}$.
The two-dimensional square lattice can be decomposed into two antiparallel sublattices A and B, according to Fig.~\ref{fig:FigureGroundStates}(a) and (b).
Fourier transformed operators are given by:
\begin{align}
    \hat{a}_{i} =& \frac{1}{\sqrt{N}}\sum_{\bm{q}} \exp(\mathrm{i} \bm{k}\cdot \bm{r}_i    ) \hat{a}_{\bm{q}}   \nonumber  \\ 
    \hat{a}^\dagger_{i} =&\frac{1}{\sqrt{N}} \sum_{\bm{q}} \exp(-\mathrm{i} \bm{k}\cdot \bm{r}_i    ) \hat{a}^\dagger_{\bm{q}}
\end{align} 
Here, $N/2$ corresponds to the number of atoms on each antiferromagnetic sublattice.

We introduce an orthonormal coordinate frame ($\hat{e}_x$, $\hat{e}_y$, $\hat{e}_z$), where $\hat{e}_z$ is parallel to the classical dipole moment $\bm{S}_{\mathrm{A}}$ on sublattice A. Raising and lowering operators are defined as $S_i^{+}=S^x+iS^y$ and $S_i^{-}=S^x-iS^y$ and fulfill the relation:
\begin{align}
  \bm{S}_i \cdot \bm{S}_j   = S_i^z S_j^z + \frac{1}{2}(S_i^+ S_j^- + S_i^- S_j^+)
\end{align}

Holstein-Primakoff bosons on sublattice A represented by $a$, $a^{\dagger}$ satisfy:
\begin{align}
    S_i^{+} &= \sqrt{2S} a_i \nonumber \\
    S_i^{-} &= \sqrt{2S} a^{\dagger}_i \nonumber \\    
    S_i^z &= S - a_i^\dagger a_i
\end{align}
We note that the expansions below can only be justified, as long as $\left\langle a_i^\dagger a_i\right\rangle\ll2\cdot S=1$.

On the second sublattice, we define modified Holstein-Primakoff operators given by:
\begin{align}
    \Tilde{S}_i^z &= -S_i^z \nonumber \\
    \Tilde{S}_i^x &= S_i^x \nonumber \\
    \Tilde{S}_i^y &= -S_i^y 
\end{align}
Holstein-Primakoff bosons therefore display equal commutation relations on both sublattices. To diagonalize the Hamiltonian, we distinguish between paths that connect sites on the same sublattice, $\mathcal{F}_A$ and $\mathcal{F}_B$, and that connect sites on different sublattices, $\mathcal{AF}$, and find:
\begin{align}
    H =& \sum_{(ij ) \in \mathcal{F}_A} J_{ij} \left[   S^z_i S^z_j   +  \frac{1}{2}(S_i^+ S_j^- + S_i^- S_j^+)   \right]   +           \nonumber \\
        &+ \sum_{(ij )\in \mathcal{F}_B} J_{ij} \left[   \tilde{S}^z_i \tilde{S}^z_j   +  \frac{1}{2}(\tilde{S}_i^+ \tilde{S}_j^- + \tilde{S}_i^- \tilde{S}_j^+)   \right]   +           \nonumber \\
        &+ \sum_{( ij ) \in \mathcal{AF}} J_{ij} \left[  - S^z_i \tilde{S}^z_j   +  \frac{1}{2}(S_i^+ \tilde{S}_j^+ + S_i^- \tilde{S}_j^-)   \right]
\end{align}

In order to write down correlation functions in closed expression, it is convenient to introduce the following sets of bonds, $\bm{\delta}_{ij}$ within the two-dimensional square lattice: $\mathcal{F}$, which contains all intra-sublattice bonds, and $\mathcal{AF}$, which contains all inter-sublattice bonds.

We define the Luttinger-Tisza exchange via:
\begin{align}
    \mu_{\bm{q}} = \sum_{\bm{\delta}_{ij}\in \mathcal{F}} J_{ij} \exp(\mathrm{i} \bm{q}\cdot\bm{\delta}_{ij})
\end{align}
and:
\begin{align}
    \gamma_{\bm{q}} = \sum_{\bm{\delta}_{ij}\in \mathcal{AF}} J_{ij} \exp(\mathrm{i} \bm{q}\cdot\bm{\delta}_{ij})
\end{align}
Eventually, we consider:
\begin{align}
    A_{\bm{q}}:=& S \left[\gamma(0) + \mu(\bm{q}) - \mu(0) -2\kappa \right]  \nonumber\\
    B_{\bm{q}}:=& S\gamma(\bm{q}) \, 
\end{align}
and write the Hamiltonian in terms of:
\begin{align}
    H =& H_0  +      \sum_{\bm{q}} a_{\bm{q}}^{\dagger} a_{\bm{q}} A_{\bm{q}}+    \nonumber \\
         &+\frac{1}{2}\sum_{\bm{q}}  B_{\bm{q}} \left[a_{\bm{q}}^{\dagger} a_{-\bm{q}}^{\dagger}      +a_{\bm{q}} a_{-\bm{q}}  \right]   \, .
\end{align}

To diagonalize the Hamiltonian, we deploy a Bogoliubov transformation, such that:
\begin{align}
    \alpha_{\bm{q}} &= \cosh \theta_{\bm{q}} a_{\bm{q}} - \sinh \theta_{\bm{q}} a_{-\bm{q}}^{\dagger}     \nonumber \\
    a_{\bm{q}} &= \cosh \theta_{\bm{q}} \alpha_{\bm{q}} + \sinh \theta_{\bm{q}} \alpha_{-\bm{q}}^{\dagger} 
\end{align}

Choosing $\theta_{\bm{q}}$ such that that :
\begin{align}
    \tanh 2\theta_{\bm{q}} = - \frac{B_{\bm{q}}}{A_{\bm{q}}}      \, .
\end{align}
the Bogoliubov coefficients take the form:
\begin{align}
    \cosh^2(2\theta_{\bm{q}}) = \frac{A^2_{\bm{q}}}{A^2_{\bm{q}}-B^2_{\bm{q}}} \\
    \sinh^2(2\theta_{\bm{q}}) = \frac{B^2_{\bm{q}}}{A^2_{\bm{q}}-B^2_{\bm{q}}}
\end{align}

And therefore:
\begin{align}
    u_{\bm{q}} = \cosh(\theta_{\bm{q}}) = \sqrt{\frac{A_{\bm{q}}+\hbar\omega_{\bm{q}}}{2\hbar\omega_{\bm{q}}}}   \\
     v_{\bm{q}} = \sinh(\theta_{\bm{q}}) = s\cdot\sqrt{\frac{A_{\bm{q}}-\hbar\omega_{\bm{q}}}{2\hbar\omega_{\bm{q}}}}    \,
\end{align}
where $s=B_{\bm{q}}/|B_{\bm{q}}|$ (see also \cite{boothroyd2020principles}).

The magnon dispersion is given by:
\begin{align}
    \hbar \omega_{\bm{q}} = \sqrt{A^2_{\bm{q}}-B^2_{\bm{q}}} 
\end{align}

and the dynamic spin structure factor by:

\begin{align}
    S_{yy}(\bm{q},\omega) =   S\cdot\frac{A_{\bm{q}}-B_{\bm{q}}}{2\hbar\omega_{\bm{q}}}
\end{align}

\subsection{Spin structure factor in the two antiferromagnetic phases}

The spin structure factor may in both antiferromagnetic phases be written as
\begin{align}
     S_{yy}(\bm{q},\omega) =: \frac{S^2}{2}\cdot \frac{Z_{\bm{q}}}{N_{\bm{q}}}
\end{align}

In the $(\pi\pi)$ antiferromagnetic limit, nominator and denominator are given by:
\begin{align}
Z_{\bm{q}}=&2J_2\cdot\cos((\bm{e}_x+\bm{e}_y)\cdot\bm{q})+2J_2\cdot\cos((\bm{e}_x-\bm{e}_y)\cdot\bm{q})-\nonumber\\&-4J_2+4J_1-2\kappa-[2J_1\cdot\cos(\bm{e}_x\cdot\bm{q})\nonumber\\ &+2J_1\cdot\cos(\bm{e}_y\cdot\bm{q}) ]
\end{align}
and
\begin{align}
N_{\bm{q}}=&\{[2J_2\cdot\cos((\bm{e}_x+\bm{e}_y)\cdot\bm{q})+2J_2\cdot\cos((\bm{e}_x-\bm{e}_y)\cdot\bm{q})-\nonumber\\&-4J_2+4J_1-2\kappa]^2-[2J_1\cdot\cos(\bm{e}_x\cdot\bm{q})\nonumber\\ &+2J_1\cdot\cos(\bm{e}_y\cdot\bm{q}) ]^2 \}^{\frac{1}{2}} 
\end{align}

In turn, for the $(\pi0)$ antiferromagnet we get:

\begin{align}
Z_{\bm{q}}=&2J_1\cdot\cos(\bm{e}_y\cdot\bm{q})+4J_2-2\kappa+\nonumber\\&-[2J_2\cdot\cos((\bm{e}_x+\bm{e}_y)\cdot\bm{q})+2J_2\cdot\cos((\bm{e}_x-\bm{e}_y)\cdot\bm{q}) + \nonumber\\ &+2J_1\cdot\cos(\bm{e}_x\cdot\bm{q})]
\end{align}

\begin{align}
N_{\bm{q}}=&\{[2J_1\cdot\cos(\bm{e}_y\cdot\bm{q})+4J_2-2\kappa]^2+\nonumber\\&-[2J_2\cdot\cos((\bm{e}_x+\bm{e}_y)\cdot\bm{q})+2J_2\cdot\cos((\bm{e}_x-\bm{e}_y)\cdot\bm{q}) + \nonumber\\ &+2J_1\cdot\cos(\bm{e}_x\cdot\bm{q}) ]^2\}^{\frac{1}{2}} 
\end{align}

\subsection{QFI for the square lattice}

We now demonstrate that the enhancement of QFI across multiple wave vectors translates to higher degree of entanglement in the semiclassical square lattice ground state.

\begin{figure}
    \includegraphics[]{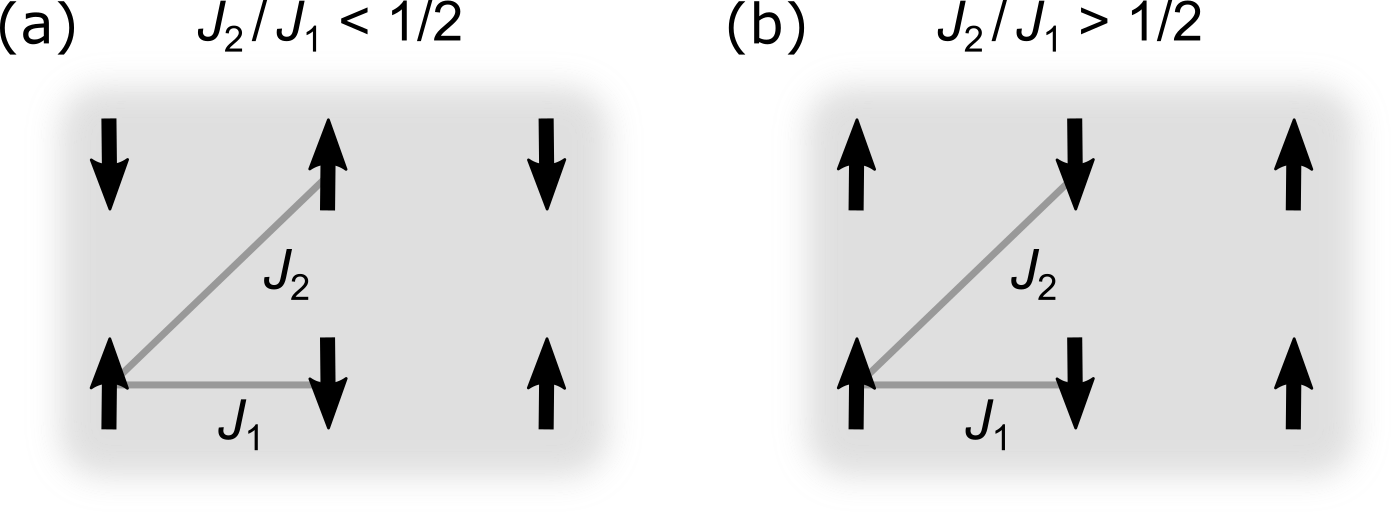} 
    \caption{Ground states of the square lattice antiferromagnetic Heisenberg model with nearest- and next-nearest neighbor exchange ($J_1$, $J_2$). (a) For $J_2/J_1\ll 1/2$, the ground state may be approximated by a $(\pi\pi)$ antiferromagnet and (b) for $J_2/J_1\gg 1/2$ by a $(\pi0)$ antiferromagnet.}
    \label{fig:FigureGroundStates}
\end{figure}

We illustrated the ground states of the $(\pi\pi)$-phase and the $(\pi0)$ phase, cf. Ref.~\cite{2009_Murg_PhysRevB}, in Fig.~\ref{fig:FigureGroundStates}(a) and (b), respectively.
However, semiclassical ground states of the AFM square lattice determined in the context of LSWT correspond only approximately to the dipole textures depicted in Fig.~\ref{fig:FigureGroundStates} and are instead characterized by finite occupation numbers of Holstein-Primakoff bosons. For a given momentum transfer $\bm{k}$, the number of bosons in the ground state is directly related to the Bogoliubov parameters and given by:
\begin{align}
     \left\langle \hat{n}_{\bm{k}}  \right\rangle =   \left\langle \hat{a}_{\bm{k}}^{\dagger}\hat{a}_{\bm{k}}  \right\rangle = \sinh^2(\theta_{\bm{k}}) = v_{\bm{k}}^2   .
\end{align}
Further, the presence of zero point fluctuations leads to a decrease of sublattice magnetization to 
\begin{align}
    M=S-\frac{1}{N} \sum_{\bm{k}} \sinh^2(\theta_{\bm{k}})  =S-\frac{1}{N} \sum_{\bm{k}}v_{\bm{k}}^2 \,.
\end{align}

We now look closer at the semiclassical ground state, which according to Refs.~\cite{2019_Kamra_PhysRevB,2022_Wuhrer_PhysRevB} can be written as a squeezed state given by: 

\begin{align}
    \ket{\Psi}_0 = \prod_{\bm{k}} S(r_{\bm{k}})\ket{\text{N\'eel}}   \, ,
\end{align}
where $\ket{\text{N\'eel}}$ corresponds to the ket state associated with the classical dipole texture. The squeezing parameter is given by $S(r_{\bm{k}})=\exp[r_{\bm{k}}(\hat{a}_{\bm{k}}\hat{a}_{\text{-}\bm{k}}+\hat{a}^{\dagger}_{\bm{k}}\hat{a}^{\dagger}_{\text{-}\bm{k}}        )]$. The AFM ground state is therefore obtained from the N\'eel state by pair-wise squeezing, which leads to quantum entanglement in the ground state~\cite{2019_Kamra_PhysRevB,2022_Wuhrer_PhysRevB}.

\begin{figure}
    \includegraphics[]{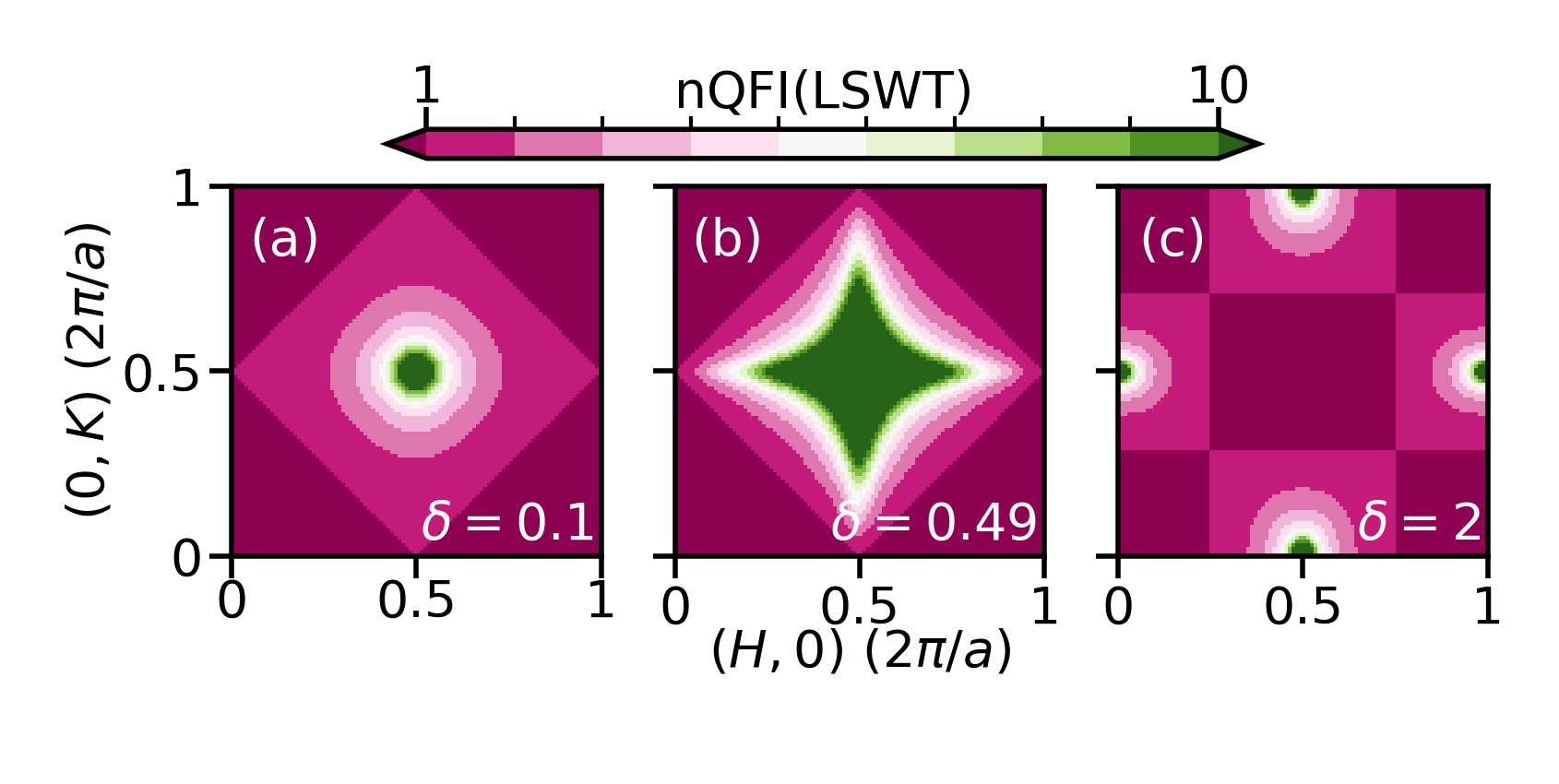} 
    \caption{Momentum-space dependence of nQFI in the square-lattice antiferromagent. Color-coding reflects nQFI as a function of $H$ and $K$, as calculated by linear-spin wave theory. Calculations were done in (a) the $(\pi\pi)$ antiferromagnetic phase for $\delta=J_2/J_1=0.1$, (b) close to the critical point at $\delta=J_2/J_1=0.49$ and (c) in the $(\pi0)$ antiferromagnetic phase at $\delta=2$.}
    \label{fig:Figure1b-squarelattice}
\end{figure}

The squeezing parameter is related to the Bogoliubov coefficients via $u_{\bm{k}}=\cosh(r_{\bm{k}})$. 
Larger values of $r$ correspond to stronger squeezing, which tends to be related to higher degree of entanglement in the resulting wave function, cf. Refs.~\cite{2019_Kamra_PhysRevB,2022_Wuhrer_PhysRevB}. As nQFI and $r_{\bm{k}}$ both increase for increasing $u_{\bm{k}}$, a larger nQFI at a given wave-vector $\bm{k}$ reflects stronger squeezing for a given $\bm{k}$-vector, therefore suggesting that also the entanglement depth for $\bm{k}$-modulated kets in the ground state is larger.

We can apply this knowledge directly to the square lattice antiferromagnet. Fig.~\ref{fig:Figure6} shows normalized quantum Fisher Information as a function of momentum transfer, calculated for the $yy$-component of the spin correlation function. (a) shows result for the $(\pi\pi)$-phase (for $\delta:=J_2/J_1=0.1$) and (c) for the $(\pi0)$-phase (for $\delta=2$). In both cases, nQFI is relatively flat throughout momentum space, except around the magnetic ordering vectors. For example, in the $(\pi0)$ phase, the reciprocal space volume $V_{\bm{k},nQFI>5}$ for that nQFI witnesses at least 6-partite entanglement occupies three percent.
Closer to the critical point (see panel (b) showing $\delta=0.49$) the portion exceeds 18 percent, therefore implying stronger squeezing for a larger number of wave-vectors.

\subsubsection*{Asymptotic behavior and scaling of nQFI}

\begin{figure}
    \includegraphics[]{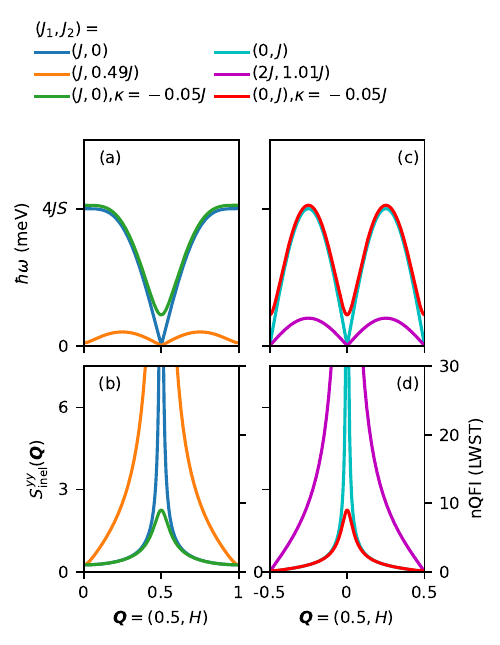} 
    \caption{Magnon dispersion and normalized Quantum Fisher information in the square-lattice antiferromagnet for different values of $J_1$, $J_2$, and magnetic anisotropy $\kappa$. Magnon dispersions are shown in (a) and (c) for the $(\pi\pi)$-phase and the $(\pi0)$-phase, respectively. Normalized Quantum Fisher information (right vertical axis) is shown in (b) and (d), respectively. As a point of reference, the left vertical axis shows the energy-integrated dynamical spin structure factor for the $yy$-component.}
    \label{fig:Figure5}
\end{figure}

Fig.~\ref{fig:Figure5} illustrates asymptotic behavior of the magnon dispersion as well as asymptotic scaling of nQFI, when approaching the critical point. 
(a) shows the magnon dispersion within the $(\pi\pi)$ phase on the momentum space line $\bm{Q}=(0.5,H)$. 
The integrated magnetic structure factor and nQFI, which differ only by a constant factor, are illustrated in (b) in same colors. Both quantities are diverging at the magnetic ordering vectors, therefore implying infinite entanglement depth. 
nQFI on this line follows the scaling ${\mathrm{nQFI}}\sim$
\begin{align}
     \frac{3J_1-2J_2-\kappa-(J1+2J_2)\cdot\cos(2\pi H)}{\sqrt{(-2J_1+2J_2+\kappa+2J_2\cos(2\pi H))^2-4J_1^2\sin^4(\pi H)}}  \, .
\end{align}

In the presence of only nearest neighbor exchange ($J_2=0$), magnons display bandwidth $4JS$.
For coupling constants $\delta$ closer to the transition and zero anisotropy ($\kappa=0$), the magnon bandwidth is gradually suppressed to zero and at the same time, nQFI diverges on the entire momentum space line. Panels (c) and (d) illustrate analogous calculations for the $(\pi0)$ phase. 

\begin{figure}
    \includegraphics[]{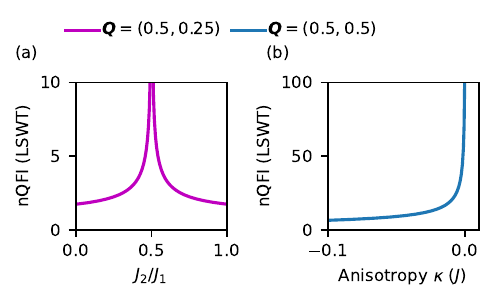} 
    \caption{(a) Variation of nQFI in the square lattice antiferromagnet as a function of $J_2/J_1$. At wave-vector $\bm{Q}=(0.5,0.25)$, nQFI diverges when approaching the critical point $1/2$. (b) Divergence of nQFI at wave-vector $\bm{Q}=(0.5,0.5)$ as a function of anisotropy in the $(\pi\pi)$-phase ($J_1=1$ and $J_2=0$).}
    \label{fig:Figure6}
\end{figure}

In turn, the divergence as a function of $J_1$ and $J_2$ for a given wave-vector $\bm{Q}=(0.5,0.25)$ ($\kappa=0$) is illustrated in Fig.~\ref{fig:Figure6}(a). In phase $(\pi\pi)$, this divergence follows the scaling
\begin{align}
    {\mathrm{nQFI}}\sim \frac{3J_1-2J_2}{\sqrt{(2J_2-2J_1)^2-J_1^2}} \, ,
\end{align}
and

Fig.\ref{fig:Figure5} shows dispersion as well as the respective quantum Fisher Information for a $(\pi\pi)$-AFM (green line) and for a $(\pi0)$-AFM (red line) with comparatively small values of anisotropy. In both cases the magnon dispersion becomes gapped at the magnetic ordering vector and the divergence of nQFI is suppressed. The decrease of nQFI as a function of anisotropy is further illustrated in Fig.~\ref{fig:Figure6}(b) for $J_2=0$ at the magnetic ordering vector.

\section{\texttt{Sunny} Simulations}
\label{Appendix2}
In the following, we present methodological details on the calculations carried out with \texttt{Sunny}.

\subsection{Triangular lattice}
The dynamical structure factor for the triangular lattice was calculated with the software package \texttt{Sunny}~\cite{2025_Dahlbom_}. Calculations were done on the smallest possible system size that reflects the translational properties of magnetic ground states. In the $120^{\circ}$ phase and stripe AFM phase these were unit cells with $N_{at}=3$ and $N_{at}=2$ number of atoms, respectively. Prior to doing linear spin-wave theory the, respective ground state was determined with the \texttt{Sunny}-internal minimization routine. We introduced an easy-plane anisotropy given by $10^{-5}\hat{S}_z^2$ in order to impose restrictions on possible domains of the randomly chosen ground state. For the calculation of structure factors, we further averaged over all orientational domains associated with $120^{\circ}$ and $240^{\circ}$ rotations around the z-axis according to hexagonal symmetry.

\subsection{Pyrochlore lattice}
Dynamical magnetic susceptibility, $S_{yy}$, was calculated in \texttt{Sunny} using a unit cell that corresponds to the crystallographic cell in Fig.~\ref{fig:Figure2}(e) and normalized to the number of atoms $N_{at}=16$. Ground states calculated by the \texttt{Sunny} internal minimization procedure are subject to randomness associated with domains in cubic symmetry. We therefore repeated the minimization at each step until the following conditions were satisfied: (i) the dipole moment $(d_1^x,d_1^y,d_1^z)$ on the position $(0.5,0,0)$ satisfies $d_1^x<0$, $d_1^y<0$ and $d_1^z$>0 and (ii) $|d_1^x-d_1^y|<0.03$. These conditions ensure that the system always results in the same domain.

\subsection{Spin ladder}
In order to account for entangled ground states in the system, we used the \texttt{Sunny}-internal entanglement mode in which neighboring spins can have a joint wave function given by
$\ket{\Psi}_{\bm{S}_1,\bm{S}_2}$ ($\mathrm{SU}(4)\rightarrow\mathbb{C}^2$). For a given external magnetic field, $\mu_0H$, ground state parameters were first randomized and subsequently determined with the \texttt{Sunny} internal minimization procedure.

\subsection{Heisenberg-Kitaev Model \label{app:Heis-Kitaev}}

\begin{figure*}
    \includegraphics[]{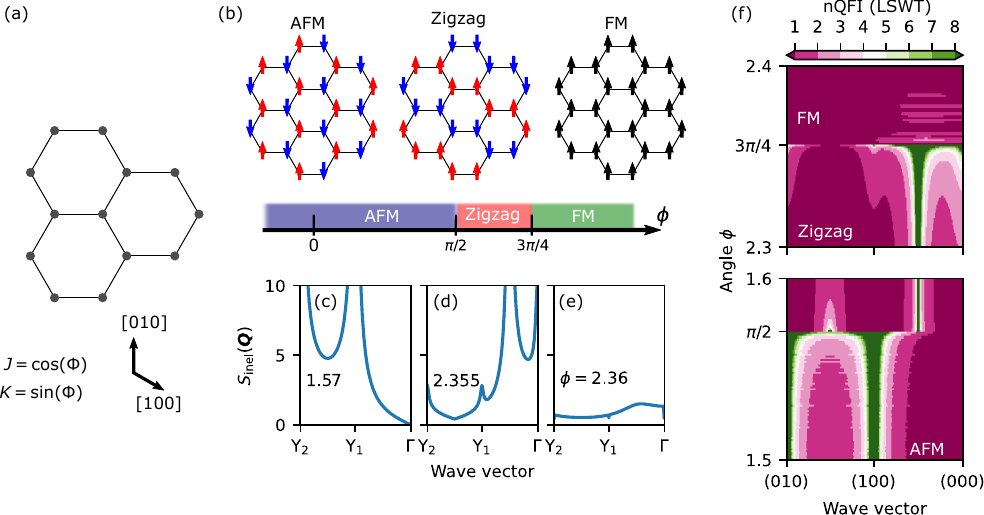} 
    \caption{Semiclassical simulations for the spin-$\frac{1}{2}$ Heisenberg-Kitaev model on the honeycomb lattice [panel (a)]. We considered a honeycomb lattice in hexagonal geometry (SG 162) with two atoms at $(\frac{2}{3},\frac{1}{3},0)$ and $(\frac{1}{3},\frac{2}{3},0)$ in a conventional hexagonal cell. Along nearest neighbor bonds, we considered isotropic Heisenberg exchange, $J$, as well as Kiteav interactions of strength $K$ that satisfy $J^2+K^2=1$. The Kitaev interactions act along an Ising axis, which is perpendicular to each nearest-neighbor bond and which encloses an angle of $\atan(\frac{1}{\sqrt{2}})$ with the basal hexagonal plane. The interactions $J$ and $K$ are parametrized by an angle $\Phi\in[0,2\pi)$. 
    (b) Classical magnetic phase diagram. We investigated the parameter range in which AFM, Zigzag, and FM phases emerge. The schematic pictures illustrate semiclassical ground states. The direction of moments is explained in the text.
    (c-e) Dynamical spin structure factor at angles $\Phi=1.57$ [panel (c)], $\Phi=2.355$ [panel (d)] and $\Phi=2.36$ [panel (e)]. The momentum-space path is defined along the points Y$_2$, Y$_1$, and $\Gamma$, which correspond to $(0,1,0)$, $(1,0,0)$, and $(0,0,0)$, respectively. 
    (f) nQFI around the transitions between phases AFM and zigzag (bottom panel) and zigzag and FM (top panel).
    }
    \label{fig:FigureS2}
\end{figure*}

We also calculated the LSWT and QFI of the $S=\frac{1}{2}$ Kitaev-Heisenberg model on the honeycomb lattice [see Fig. \ref{fig:FigureS2}(a)]. 
We consider Heisenberg interactions $J=\cos(\phi)$ and Kitaev interactions $K=\cos(\phi)$ parametrized by an angle $0\leq\phi<2\pi$. The Ising direction is chosen perpendicular to each bond enclosing an angle $\atan(\frac{1}{\sqrt{2}})$ with the basal hexagonal plane (which preserves the three-fold $c$-axis rotation symmetry of the honeycomb lattice).

The classical phase diagram as a function of $\phi$ is presented in Fig. \ref{fig:FigureS2}(b), comprising AFM, zigzag, and FM phases, subsequently accessed as a function of increasing $\phi$~\cite{2014_Rau_PhysRevLett}\footnote{We do not consider the stripy phase for $\frac{3\pi}{2}\leq \phi \leq \frac{7\pi}{4}$ in this work.}. The spin structures are shown in the same panel. In the AFM and FM phases, magnetic moments may point along any direction due to an accidental spin rotation symmetry~\cite{2014_Rau_PhysRevLett}. In the zigzag phase, the moments are pinned along the Ising axis. For simplification, moments in (b) are drawn along $[010]$.

Dynamical magnetic susceptibility, $S_{\mathrm{inel}}(\bm{Q})=S^{xx}_{\mathrm{inel}}(\bm{Q})+S^{yy}_{\mathrm{inel}}(\bm{Q})+S^{zz}_{\mathrm{inel}}(\bm{Q})$, for different values of $\phi$ is presented in Fig. \ref{fig:FigureS2}(c)-(e). While the angles $\phi=1.57$ and $2.355$ correspond to points in the antiferromagnetic phases close to quantum phase transitions, $\phi=2.36$ corresponds to the ferromagnetic phase.

nQFI, obtained by normalizing $S_{\mathrm{inel}}(\bm{Q})$ by $4/12S^2$, is presented in Fig. \ref{fig:FigureS2}(f). At $\phi=\frac{\pi}{2}$, the transition between phases AFM and Zigzag, nQFI diverges on the entire line connecting $(010)$ and $(100)$. Semiclassical calculations therefore herald a highly entangled ground state, in agreement with quantum phase diagrams, which predict a Kitaev spin liquid phase in this area, cf. Ref.~\cite{2014_Rau_PhysRevLett,2013_Chaloupka_PhysRevLett}. It is worth noting that the dynamical magnetic susceptibility at $\phi=\pi/2$ can be calculated exactly \cite{2014_Knolle_Kitaev}, yielding a spectrum with finite spin gap, unlike the LSWT prediction [e.g., Fig. \ref{fig:FigureS2}(c)].

Interestingly, the Kitaev-only point is not the only place where LSWT nQFI diverges along a line of reciprocal space.  nQFI is also strongly enhanced across multiple wave vectors at $\phi=\frac{3\pi}{4}$, where numerical simulations have shown the magnetic order vanishes at the same point~\cite{2019_Rusnacko_PhysRevB, 2024_Georgiou_PRR}; this suggests that perhaps the boundary between FM and zigzag order also hosts an emergent quantum phase. 
If indeed this were the case, this would be a strong validation that QFI can witness emergent physics. We also note that a recent numerical study on a two-leg ladder geometry found nontrivial frustration effects at the same quantum phase transition, but did not identify an emergent phase \cite{Pandey2026KHladder}.

\section{Dependence of semiclassical simulations on the spin operator component}
\label{Appendix3}
\begin{figure}
	\includegraphics[]{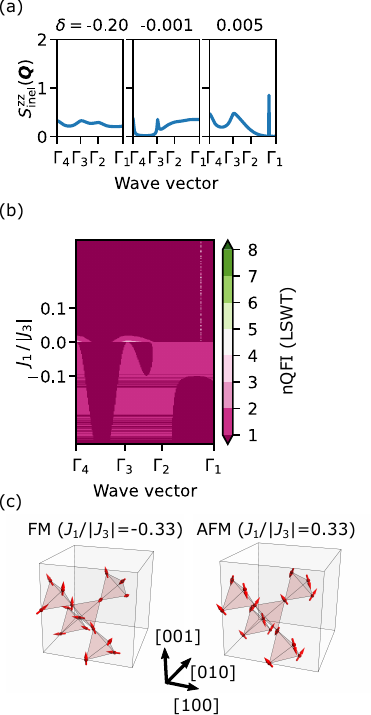}
	\caption{\label{fig:FigureS4}(a) Energy integrated spin structure factor calculated with \texttt{Sunny} in the same way as in Fig.~\ref{fig:Figure2}(k), but for the $zz$-spin component of the pyrochlore lattice. (b) nQFI inferred from the $zz$-component of the spin structure factor of the pyrochlore lattice. (c) Spin textures in phases FM and AFM in one crystallographic unit cell.}
\end{figure}

To illustrate the dependence of our results on the component of the spin structure factor, we repeated the calculations on the pyrochlore lattice shown in Fig.~\ref{fig:Figure2} (k) and (l), but considered the $zz$-spin structure component, instead of the $yy$-component. The results are presented in Fig.~\ref{fig:FigureS4}. The result for the two spin components are distinctively different, reflecting that the two operators have different variances.

\section{Scaling in the one-dimensional spin chain}
\label{Appendix4}
\subsubsection{Methods}
We consider the antiferromagnetic Heisenberg chain
\begin{align}
H	&=	\sum_{i=1}^L \bm{S}_i \cdot \bm{S}_{i+1},
\end{align}
where $\bm{S}_i = \bm{\sigma}_i/2$ is the spin-1/2 operator, and $L$ is the chain length. We exploit that the quantum Fisher information (QFI) density related to the $S^z_Q$ operator can, in the $T\rightarrow 0$ limit, be expressed
\begin{align}
f_\mathcal{Q}(Q)	&=	4S(Q) = \frac{4}{L} \left[ \langle \mathcal{O}^2\rangle - \langle \mathcal{O}\rangle^2 \right],
\end{align}
where $S(Q)$ is the static spin structure factor (cf. Ref.~\cite{2023_Menon_PRB}). With $\mathcal{O}=\mathcal{O}_{Q}=\sum_j S_j^{z}\exp(\mathrm{i}Qj)$ we can evaluate nQFI using Eq. \eqref{eq:nQFI}. Because of the SU(2) symmetry of the Heisenberg Hamiltonian, all three spin components are equivalent. By calculating nQFI from $S(Q)$ evaluated in the ground state we obtain theoretically maximal values, not subject to the same energy-filtering or resolution effects that affect nQFI calculated as an integral over the dynamical susceptibility. We focus on the antiferromagnetic wave vector $Q=\pi$, which maximizes the nQFI.

We calculate $S(Q)$ (and hence nQFI) using Lanczos exact diagonalization \cite{1950_Lanczos} as implemented in the H$\Phi$ library \cite{2017_Kawamura_ComputerPhysicsCommunications, 2024_Ido_ComputerPhysicsCommunications} and the density-matrix renormalization group (DMRG) \cite{1992_White_PhysRevLett, 1993_White_PhysRevB} as implemented in the DMRG++ software \cite{2009_Alvarez_ComputerPhysicsCommunications}. Results were obtained for both periodic boundary conditions (PBC) and open boundary conditions (OBC). Lanczos was used for system sizes $2\leq L \leq 36$ with both boundary conditions, and DMRG for system sizes $4\leq L \leq 1024$ with OBC ($4\leq L \leq 256$ with PBC). Truncation errors smaller than $10^{-12}$ ($10^{-8}$) were achieved in the OBC (PBC) DMRG calculations by keeping less than $1000$ ($2000$) U(1) states. In addition, PBC results for short chains ($L=2$ and $L=4$) were verified through direct diagonalization. 

\subsubsection{Results}

\begin{figure}
	\includegraphics[width=0.45\textwidth]{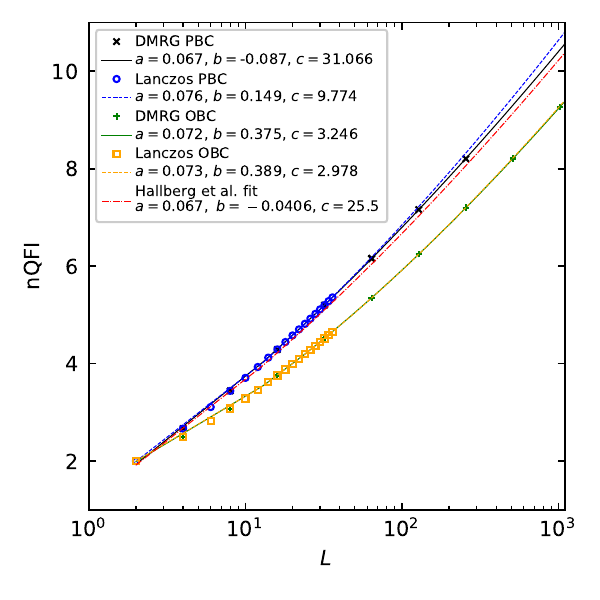}
	\caption{\label{fig:QFIscaling}Finite-size scaling of the normalized QFI for the isotropic spin-1/2 antiferromagnetic Heisenberg chain at zero temperature. The upper (lower) set of curves correspond to PBC (OBC). The open blue circles represent the PBC Lanczos results. Black crosses represent the DMRG PBC results. Fits to these lines agree well, but slightly diverge at large system size. For comparison, the original fit obtained by Hallberg et al. for $L\leq 70$ \cite{1995_Hallberg_PhysRevB} is shown as as a dot-dashed red line. For the OBC curves, orange open squares indicate Lanczos results and green plus signs indicate DMRG results. Fits to these data agree across well across all studied system sizes.}
\end{figure}

Our numerical results are shown in Fig.~\ref{fig:QFIscaling}. We found that individual curves can be well fit to power laws with an offset, i.e. $f_\mathcal{Q}(Q=\pi) = a L^k +b$. However, the resulting fits against Lanczos or DMRG data diverge at large system sizes. We thus instead fit the data to
\begin{equation}
\mathrm{nQFI}(Q=\pi)	= 4\left[ b + \frac{4}{3} a \left[ \ln \left( \frac{cL}{2}\right)\right]^{3/2}\right],
\end{equation}
where the expression inside the brackets is an expression Hallberg et al. \cite{1995_Hallberg_PhysRevB} found fits $S(Q=\pi)$ well. Using this form, we find extremely close agreement between Lanczos and DMRG for OBC over all system sizes studied. For PBC we also see a close agreement, but the fits begin to diverge somewhat for chain lengths $L>100$. This may be due to higher truncation errors in the DMRG PBC calculations, or a different approach to the thermodynamic limit in the Lanczos data. The figure also shows Hallberg et al.'s original fit, obtained using DMRG data from rings up to $L=70$. Regardless of boundary conditions, the nQFI value at $L=2$ is found to equal $2$, consistent with with the value found for isolated dimers \cite{scheie2025tutorial}. These findings establish that the asymptotic scaling is well described by Eq.\eqref{eq:HAFM:scalingform}.

\clearpage
\pagebreak

\end{document}